\documentclass[twocolumn,preprintnumbers,amsmath,amssymb]{revtex4}
\usepackage{amsmath,amsfonts,latexsym,amssymb,graphicx,graphics,epsfig,subfigure,color,makeidx}
\usepackage{xcolor,diagbox}
\usepackage{multirow}
\usepackage[colorlinks,linkcolor=blue,anchorcolor=blue,citecolor=green,urlcolor=blue]{hyperref}
\usepackage{mathrsfs}
\usepackage{amsmath}
\usepackage{multirow}
\usepackage{array}
\newcommand{\PreserveBackslash}[1]{\let\temp=\\#1\let\\=\temp}
\newcolumntype{C}[1]{>{\PreserveBackslash\centering}p{#1}}
\newcolumntype{R}[1]{>{\PreserveBackslash\raggedleft}p{#1}}
\newcolumntype{L}[1]{>{\PreserveBackslash\raggedright}p{#1}}

\begin{document}
		\renewcommand{\baselinestretch}{1.3}
		\newcommand\beq{\begin{equation}}
			\newcommand\eeq{\end{equation}}
		\newcommand\beqn{\begin{eqnarray}}
			\newcommand\eeqn{\end{eqnarray}}
		\newcommand\nn{\nonumber}
		\newcommand\fc{\frac}
		\newcommand\lt{\left}
		\newcommand\rt{\right}
		\newcommand\pt{\partial}
		\allowdisplaybreaks
	
	\title{Motion of spinning particles around black hole in a dark matter halo}

	\author{Qin Tan$^{a}$}
	\author{Weike Deng$^{a}$}
	\author{Sheng Long$^{a}$}
	\author{Jiliang Jing$^{a}$}\email{jljing@hunnu.edu.cn, corresponding author.}
	
	\affiliation{$^{a}$Department of Physics, Key Laboratory of Low Dimensional Quantum Structures and Quantum Control of Ministry of Education, Synergetic Innovation Center for Quantum Effects and Applications, Hunan Normal University, Changsha, 410081, Hunan, China}

\begin{abstract}

The motion of a rapidly rotating object in curved spacetime is affected by the spin-curvature force, an effect captured in the motion of spinning test particles. Recently, Cardoso et al.~[Phys. Rev. D 105, L061501 (2022)] found an exact solution describing a black hole immersed in a Hernquist distribution of dark matter. In this work, we investigate the motion of spinning particles around this black hole. We use the Mathison-Papapetrou-Dixon equation and the Tulczyjew spin-supplementary condition to calculate the effective potential, four-momentum, and four-velocity of the spinning particle. The equatorial motion of spinning test particles and the properties of the marginally bound orbits, innermost stable circular orbits, and periodic orbits are further studied. We find that the existence of dark matter halos can significantly change the orbital eccentricity, energy, and the marginally bound orbits, innermost stable circular orbits, and periodic orbits parameters of spinning test particles. Compared to the Schwarzschild black hole, dark matter halos bring the marginally bound orbit and innermost stable circular orbit of a spinning test particle closer to the event horizon. These results could help us understand the properties of black holes in dark matter halos.

\end{abstract}

\maketitle

\section{Introduction}
Dark matter is one of the most fascinating mysteries in modern astronomy. Through astrophysical and cosmological observations at various scales, such as large-scale structures~\cite{Dietrich:2012}, cosmic microwave background anisotropies~\cite{Planck:2018vyg}, and the rotation curves~\cite{Begeman:1991iy}, there is overwhelming evidence to support the existence of dark matter. The abundant presence of dark matter in the universe may surround galaxies and be close to black holes~\cite{Sadeghian:2013laa}, affecting the dynamics of compact binaries and the way gravitational waves propagate~\cite{Barack:2018yly,Bertone:2018krk}. Due to the boom in gravitational wave astronomy~\cite{LIGOScientific:2016aoc,LIGOScientific:2020ibl} and the construction of high-precision observatories such as the Event Horizon Telescope~\cite{EventHorizonTelescope:2019dse}, it is becoming possible to understand the nature of dark matter by using black holes that dance with dark matter.

Recently, Cardoso et al.~\cite{Cardoso:2021wlq} proposed a spherically symmetric, static, non-vacuum black hole solution. This solution represents a black hole immersed in a dark matter halo described by the Hernquist profile~\cite{Hernquist:1990}. Subsequently, the solution quickly generated a lot of interest, and some properties of the black hole were investigated, such as quasinormal modes~\cite{Konoplya:2021ube}, shadow~\cite{Xavier:2023exm}, Love numbers~\cite{Liu:2022lrg,LimaJunior:2022gko}, gravitational wave emission~\cite{Cardoso:2022whc,Figueiredo:2023gas}, solutions to other dark matter distributions~\cite{Jusufi:2022jxu,Konoplya:2022hbl,Shen:2024qxv}, the charged or rotating solutions~\cite{Stelea:2023yqo,Zhang:2024hjr}, and so on. In this paper, we study the motion of spinning test particles around this black hole. 

It is a very effective method to study the nature of a black hole by the motion of test particles near them. Changes in the background spacetime will result in changes in the motion state of the test particles. These changes will be reflected in the black hole shadow, accretion disk and so on. In previous studies, the effect of dark matter halos on the shadow of black holes has been studied~\cite{Xavier:2023exm}. However, in addition to photons, massive spinning particles or spinning stars also orbit black holes. In particular, the inspirals of stellar mass compact objects on the supermassive black hole will be detected by the space-based Observatory Laser Interferometer space antenna~\cite{Babak:2017tow}. This will allow us to accurately test the properties of supermassive objects at the center of galaxies and their surroundings. In the subleading corrections of dynamics of these so-called extreme-mass-ratio inspirals, the spin of smaller companion plays a crucial role~\cite{Huerta:2011zi,Piovano:2021iwv}. The effect of spin of smaller companion can be described by the motion of a spinning test particle around a massive compact object~\cite{Warburton:2017sxk,Skoupy:2021asz,Mathews:2021rod,Piovano:2020zin,Drummond:2023wqc}. Thus, considering the dynamics of non-zero spinning particles around a black hole may provide insight into the motion properties of stars orbiting supermassive black holes. For a spinning test particle, its motion no longer follows a geodesic due to the spin curvature force~\cite{Wald:1972sz,Hanson:1974qy}. In this case, based on the pole-dipole approximation of the spinning test particle, we can describe the motion of a rotating test particle moving against a curved background through the Mathison-Papapetrou-Dixon (MPD) equation~\cite{Mathisson:1937ms,Papapetrou:1951gr,Corinaldesi:1951gr,Tulczyjew:1959tr,Dixon:1964gr}. Previous literature have investigated the properties of spinning test particles in different black hole or other backgrounds~\cite{Suzuki:1998vy,Han:2008zzf,Jefremov:2015gza,Harms:2016ctx,LukesGerakopoulos:2017cru,Zhang:2017nhl,Mukherjee:2018zug,Zhang:2018omr,Zhang:2019oet,Antoniou:2019awm,Nucamendi:2019qsn,Zhang:2020qam,Semerak:2015dza,Mukherjee:2018dmm,Toshmatov:2019bda,Benavides-Gallego:2021lqn,Yang:2022jno,Zhang:2022qzw,Chen:2024sbc,Witzany:2023bmq,Liu:2024lda}. In this paper, we focus on the motion of spinning test particles in the black hole background with dark matter halo, and study the influence of dark matter halo on the orbital eccentricity and velocity of spinning test particles, especially the marginally bound orbits (MBO), the innermost stable circular orbit (ISCO), and the periodic orbit properties. 

The remainder of this paper is organized as follows. In Sec.~\ref{sec_solution}, we review the solution and the properties for a Schwarzschild black hole surrounded by a dark matter halo. Based on this solution, in Sec.~\ref{Sec_Motion}, we use the MPD equation and the Tulczyjew spin-supplementary condition to obtain the effective potential, four-momentum, and four-velocity of the spinning test particle. The equatorial motion of spinning test particles and the properties of the MBO, ISCO, and periodic orbit are further studied. Finally, some conclusions and discussions are provided in Sec.~\ref{sec:conclusions}.

\section{The geometry of a black hole with dark matter halo}
\label{sec_solution}

In this section, we will review the solution for a Schwarzschild black hole surrounded by a dark matter halo. To get an accurate solution for a black hole surrounded by a dark matter halo, Cardoso et al.~\cite{Cardoso:2021wlq} used a Hernquist-type density distribution to model the galactic profiles
\begin{equation}
\rho(r)=\frac{Ma_0}{2\pi\,r(r + a_0)^3}.\label{hernquist_density}
\end{equation}
Here $M$ is the total mass of the dark matter halo and $a_0$ is a typical length-scale of the galaxy. Considering that the dark matter halo has the following energy-momentum tensor
\begin{equation}
	T^{\mu}_{\nu}=\text{diag}(-\rho,0,P_{t},P_{t}),\label{energy_momentum}
\end{equation}
and the geometry has the following form
\begin{equation}
ds^2=-f(r) dt^2+\frac{dr^2}{1-2m(r)/r}+r^2d\Omega^2.\label{metric}
\end{equation}
The function $f(r)$ can be obtained by the mass function $m(r)$ as 
\begin{eqnarray}
	\frac{rf'(r)}{2f(r)}=\frac{m(r)}{r-2m(r)},\label{ffunction1}
\end{eqnarray}
The mass function $m(r)$  is chosen as the following form that is compatible with the Hernquist profile~\eqref{hernquist_density}
\begin{equation}
m(r)=M_{\rm BH}+\frac{M r^2}{(a_0+r)^2}\left(1-\frac{2M_{\rm BH}}{r}\right)^2.\label{massfunction}
\end{equation}
The above mass function describes a black hole with a mass of $M_{\rm BH}$ at small distances and corresponds to the Hernquist profile~\eqref{hernquist_density} at large scales. Combining Eqs.~\eqref{ffunction1} and \eqref{massfunction}, and imposing the asymptotic flatness condition ($f(r)\rightarrow1$ at large $r$), one obtains the specific expression of the function $f(r)$ as
\begin{eqnarray}
f(r)&=&\left(1-\frac{2M_{\rm BH}}{r}\right)e^\Upsilon,\label{ffunction2}\\
\Upsilon&=&-\pi\sqrt{\frac{M}{\xi}}+2\sqrt{\frac{M}{\xi}}\arctan{\frac{r+a_0-M}{\sqrt{M\xi}}},\\
\xi&=&2a_0-M+4M_{\rm BH}.
\end{eqnarray}
The density of matter corresponding to the above solution is
\begin{equation}
4\pi \rho(r)=\frac{m'(r)}{r^2}=\frac{2M(a_0+2M_{\rm BH})(1-2M_{\rm BH}/r)}{r(r+a_0)^3}.\label{hernquist_density2}
\end{equation}
The event horizon of this black hole is located then at $r=2M_{\rm BH}$. The ADM mass of this spacetime is $M+M_{\text{BH}}$. It is worth noting that when $M>2(a_{0}+2M_{\text{BH}})$, the Ricci scalar diverges at $r=M-a_{0}\pm\sqrt{M^{2}-2Ma_{0}-4MM_{\text{BH}}}$, that is, there are also spacetime singularities at these two places. These singularities do not affect our study, as we focus solely on the case where $M<2(a_{0}+2M_{\text{BH}})$. In order to better explain the influence of dark matter halo, we follow Ref.~\cite{Xavier:2023exm} and define the compactness $\mathcal{C}$ of dark matter halo as
\begin{equation}
	\mathcal{C}=\frac{M}{a_{0}}.\label{compactness}
\end{equation}
We will discuss in the next section how black holes with dark matter halos of different compactness affect the motion of spinning test particles.

\section{Motion of Spinning Test Particles around black hole in a dark matter halo}
\label{Sec_Motion}

\subsection{MPD Equation and Equatorial Motion}
The motion of a spinning test particle is no longer described by the geodesic equation, but by the MPD equations
\beqn
\fc{D P^{\mu}}{D \lambda}&=&-\fc{1}{2}R^\mu_{\nu\alpha\beta}u^\nu S^{\alpha\beta}, \label{EoM1}\\
\fc{D S^{\mu\nu}}{D \lambda}&=&P^\mu u^\nu-P^\nu u^\mu, \label{EoM2}
\eeqn
where $S^{\mu\nu}$ is the spin tensor associated with the particle's internal angular momentum, $P^{\mu}$, $u^\mu\equiv dx^\mu/d\tau$, and $\lambda$ are the four-momentum, the four-velocity along the trajectory, and the affine parameter, respectively. A simple count of the degrees of freedom in variables $S^{\mu\nu}$, $P^\mu$, and $u^\mu$ in the equations of motion (\ref{EoM1}) and (\ref{EoM2}) yields 13, while we have only 10 equations available. This insufficiency in solving for all variables arises from the fact that the center of mass of a spinning body in relativity is observer-dependent. Consequently, an extra ``spin-supplementary condition" must be imposed to obtain a deterministic system. Various choices for this condition exist in the literature~\cite{Costa:2014nta}. 
Due to the Tulczyjew spin-supplementary condition's~\cite{Tulczyjew:1959tr} ability to solve for the ISCO over a larger range of particle spins~\cite{Harms:2016ctx,LukesGerakopoulos:2017cru} and its better agreement with the results of Hamiltonian dynamics~\cite{Harms:2016ctx}, we adopt the Tulczyjew spin-supplementary condition in this paper. The expression for the Tulczyjew spin supplementary condition is~\cite{Tulczyjew:1959tr}
\beq
P_{\mu}S^{\mu\nu}=0. \label{Tulczyjew_condition}
\eeq
Combining the equations of motion (\ref{EoM1}), (\ref{EoM2}), and the Tulczyjew condition~\eqref{Tulczyjew_condition}, it can be obtained that the mass $m$ and spin $s$ of the particle as
\beqn
m^2&=&-P^\mu P_\mu,\label{mass}\\
s^2&=&\frac{1}{2}S^{\mu\nu}S_{\mu\nu}.\label{spin}
\eeqn

Here, we focus on the equatorial motion of spinning test particles with spin-aligned or anti-aligned orbits in the background of the black hole described by Eq. \eqref{metric}. In this case, the four-momentum and spin tensor satisfy $P^\theta=0$ and $S^{\mu\theta}=0$. Applying the Tulczyjew condition (\ref{Tulczyjew_condition}), we obtain the six non-vanishing components of the spin tensor as
\beqn
S^{r\phi} &=&-\frac{s}{r}\frac{P_t}{m\psi}=-S^{\phi r} ,\label{ex_srphi}\\
S^{\phi t}&=&-\frac{s}{r}\frac{P_r}{m\psi}=-S^{t\phi},\label{ex_sphit}\\
S^{tr}&=&-\frac{s}{r}\frac{P_\phi}{m\psi}=-S^{rt},\label{ex_str}
\eeqn
where $\psi=\sqrt{\frac{rf(r)}{r-2m(r)}}$.

Next we construct conserved quantities using the Killing vectors in this spacetime. The static spherically symmetric spacetime given by Eq. \eqref{metric} possesses a timelike Killing vector $\xi^\mu=(\pt_t)^\mu$ and a spacelike Killing vector $\eta^\mu=(\pt\phi)^\mu$. The relation between the conserved quantity $\mathcal{Q}$ and the killing vector $\mathcal{K}^\mu$ is
\beq
\mathcal{Q}=\mathcal{K}^\mu P_\mu-\frac{1}{2}S^{\mu\nu}\mathcal{K}_{\mu;\nu}.
\eeq
Consequently, two important conserved quantities emerge: the energy $e$ and the total angular momentum $j$:
\beqn
e&=&-\xi^\mu P_\mu+\frac{1}{2}S^{\mu\nu}\xi_{\mu;\nu}=-P_t-\fc{1}{2}\fc{s}{r}\fc{P_\phi}{m\psi}g'_{tt},\label{CQ_Energy}\\
j&=&\eta^\mu P_\mu -\fc{1}{2}S^{\mu\nu}\eta_{\mu;\nu}=P_\phi-\fc{1}{2}\fc{s}{r}\fc{P_t}{m\psi}g'_{\phi\phi},\label{CQ_angular_momentum}
\eeqn
where the prime denotes differentiation with respect to the radial coordinate $r$.
The non-zero components of the momentum can be derived from these equations, together with Eqs. \eqref{metric} and \eqref{mass}:
\beqn
P_t&=&- \frac{2m\psi \left( 2\psi \bar{e} + M_{\text{BH}}\bar{s} \bar{j} \partial_r g_{tt} \right)}{4 \psi^2 + 2r M_{\text{BH}}\bar{s}^2 \partial_r g_{tt}},\\
P_\phi &=& \frac{2m\psi  M_{\text{BH}}\left( 2\psi \bar{j} + 2r\bar{s} \bar{e} \right)}{4 \psi^2 + 2r M_{\text{BH}}^{2}\bar{s}^2 \partial_r g_{tt}},\\
(P^r)^2&=&- \frac{m^2 + g^{tt} (P_t)^2 + r^2(P_\phi)^2}{g^{rr}},
\eeqn
where the parameters of energy, total angular momentum, and spin have been rescaled to be dimensionless: $\bar{e}=\fc{e}{m}$, $\bar{j}=\fc{j}{mM_{\text{BH}}}$, and $\bar{s}=\fc{s}{mM_{\text{BH}}}$.

On the other hand, the equation of motion \eqref{EoM2} yields
\beqn
\fc{D S^{tr}}{D\lambda}&=&P^t u^r-P^r,\\
\fc{D S^{t\phi}}{D\lambda}&=&P^t u^\phi-P^\phi.
\eeqn
Substituting Eqs.~\eqref{ex_sphit} and~\eqref{ex_str} into these equations leads to
\beqn
P^t u^r-P^r \!&\!=\!&\!-\fc{M_{\text{BH}}}{r\psi}\fc{DP_\phi}{D\lambda}+\fc{M_{\text{BH}}P_\phi}{r \psi}\lt(\fc{1}{r}+\fc{\psi'}{\psi} \rt)  u^r,\label{RealtionI}\nonumber\\  \\ 
P^t u^\phi-P^\phi \!&\!=\!&\!\fc{M_{\text{BH}}}{r\psi}\fc{DP_r}{D\lambda}-\fc{M_{\text{BH}}P_r}{ r\psi}\lt(\fc{1}{r}+\fc{\psi'}{\psi} \rt) u^r.\label{RealtionII}
\eeqn
Moreover, using the equation of motion \eqref{EoM1}, we obtain
\beqn
\!\!\!\!\!\!\! P^t u^r \!-\! P^r \!&\!\!=\!\!&\!\fc{M_{\text{BH}}}{2 r\psi}R_{\phi\nu\alpha\beta}u^\nu S^{\alpha\beta}\!+\!\fc{M_{\text{BH}}P_\phi}{r \psi}\lt(\fc{1}{r} \!+\! \fc{\psi'}{\psi} \rt)\!  u^r,\nonumber\\  \\ 
\!\!\!\!\!\!\! P^\phi \!-\!  P^t u^\phi  \!&\!\!=\!\!&\! \fc{M_{\text{BH}}}{2 r\psi}R_{r\nu\alpha\beta}u^\nu S^{\alpha\beta} \!+\! \fc{M_{\text{BH}}P_r}{ r\psi}\lt(\fc{1}{r}\!+\!\fc{\psi'}{\psi} \rt)\! u^r.\nonumber \\ 
\eeqn
The four-velocity components can then be derived algebraically:
\beqn
u^r&=&\fc{c_1}{a_1},\label{ur}\\
u^\phi&=&\fc{a_1 c_2-a_2 c_1}{a_1 b_1 },\label{uphi}
\eeqn
where the functions $a_1$, $b_1$, $c_1$, $a_2$, $b_2$, and $c_2$ are defined as follows:
\beqn
a_1&=&P^t-\frac{M_{\text{BH}} \bar{s}}{r \psi } \left[R_{\phi rr\phi } S^{r\phi }+\left(\frac{1}{r}+\frac{\psi '}{\psi }\right)P_{\phi } \right],\\
b_1&=&P^t+\fc{M_{\text{BH}}}{r\psi}R_{r\phi r\phi}S^{r\phi},\\
c_1&=&P^r+\fc{M_{\text{BH}}}{r\psi}R_{\phi t t\phi}S^{t\phi},\label{Para_c1}\\
a_2&=& \frac{M_{\text{BH}} \bar{s} }{r \psi } \lt(\frac{1}{r}+\frac{\psi '}{\psi }\rt)P_r, \\
c_2&=&P^\phi-\fc{M_{\text{BH}}}{r\psi}R_{r t tr}S^{tr}.
\eeqn

Equations~\eqref{ex_sphit}, \eqref{ur}, and \eqref{Para_c1} reveal that $u^r \propto P_r $. Thus, even though the four-velocity $u^\mu$ is generally not parallel to the four-momentum $P^\mu$ under the Tulczyjew condition, the radial velocity $u^r$ is always parallel to the radial momentum $P^r$, which facilitates the derivation of the effective potential for spinning test particles. To simplify the calculations, instead of normalizing the four-velocity as $u_\mu v^\mu=-1$, we fix $u^t=1$ by setting the affine parameter $\lambda$ equal to the coordinate time $t$, since the trajectories of test particles are independent of the affine parameterization. Consequently, the velocities $u^r=\dot r$ and $u^\phi=\dot\phi$.

It is worth noting that, the equations of motion (\ref{EoM1}) and (\ref{EoM2}) along with the Tulczyjew condition (\ref{Tulczyjew_condition}) lead to the relation
\beq
u^\mu+ \frac{P^\nu u_\nu}{m^2}P^\mu =\fc{S^{\mu\nu}S^{\alpha\beta}R_{\nu\rho\alpha\beta}u^\rho } {2m^2}.
\eeq
This implies that, under the Tulczyjew condition, the four-velocity $u^\mu$ is generally not parallel to the four-momentum $P^\mu$. Consequently, the four-velocity $u^\mu$ may transition from timelike to spacelike, rendering the orbit unphysical. To ensure the physical validity of spinning test particles' motion, the superluminal constraint should be imposed:
\beq
\fc{u^\mu u_\mu}{(u^t)^2}=g_{tt}+g_{rr} \dot r ^2+g_{\phi\phi} \dot \phi ^2+2g_{t\phi}\dot\phi<0,
\label{Superluminal_Constraint}
\eeq
where the dot signifies the derivative with respect to coordinate time $t$. Furthermore, the superluminal issue for spinning particles within the``pole-dipole" approximation may be circumvented by considering a non-minimal spin-gravity interaction~\cite{Deriglazov:2015wde,Deriglazov:2017jub}.

\subsection{Effective potential}

The effective potential method is a powerful tool for studying the motion of a classical test particle in a central field. By obtaining the effective potential of a spinning test particle in a black hole background, we can gain insights into its motion properties. Since the radial velocity $u^r$ is proportional to the radial momentum $P^r$, we can determine the radial turning points of the particle by finding the zeros of $P^r$ instead of those of $u^r$. This allows us to derive the effective potential more conveniently using the radial momentum $P^r$. Factoring out the energy dependence from the radial dependent part of $(P^r)^2$ leads to
\beqn
\fc{(P^r)^2}{m^2}&=&A \bar{e}^2+B\bar{e}+C\nonumber\\
&=&A\left(\bar{e}-\fc{-B+\sqrt{B^2-4AC}}{2A}  \right)\nonumber\\
&&\times\left(\bar{e}-\fc{-B-\sqrt{B^2-4AC}}{2A}  \right),
\label{Motion_Eq}
\eeqn
where for the black hole described by Eq. \eqref{metric}, the parameters $A$, $B$, and $C$ are given by
\beqn
A&=&\frac{4 \left(r^2-M_{\text{BH}}^2 \bar{s}^2 f\right)}{\left[\left(2 r-M_{\text{BH}}^2  \bar{s}^2 f'\right)\psi -2 M_{\text{BH}}^2  \bar{s}^2 f \psi'\right]^2},\\
B&=&\frac{4  \bar{s} \bar{j} M_{\text{BH}}^2  \left[2  \left(\psi -r \psi'\right)f-r \psi f'\right]}{\left[\left(2 r-M_{\text{BH}}^2  \bar{s}^2 f'\right)\psi -2 M_{\text{BH}}^2  \bar{s}^2 f \psi'\right]^2},\\
C&=&\frac{8 M_{\text{BH}}^2 \bar{s}^2 r  f \psi \psi'-4 \left[r^2+M_{\text{BH}}^2 \bar{s}^2 \left(f-r f'\right)\right]\psi^2 }{\left[(2 r-M_{\text{BH}}^2 \bar{s}^2 f')\psi-2 M_{\text{BH}}^2 \bar{s}^2 f \psi '\right]^2}\fc{\bar{j}^2}{\bar{s}^2 }\nn\\
&& +\fc{\bar{j}^2}{\bar{s}^2 }-f.
\eeqn

The effective potential of the spinning test particle is given by the positive square root of Eq.~\eqref{Motion_Eq}, as the positive square root corresponds to the four-momentum pointing toward the future, while the negative square root corresponds to the four-momentum pointing toward the past. Thus, the effective potential is expressed as
\beqn
\!\!\!\!\!\!\!\! V_{\text{eff}} \!&\!\!=\!\!&\! \fc{-B+\sqrt{B^2-4AC}}{2A}.
\label{Effective_potential}
\eeqn
It allows us to study the motion of spinning test particles in the black hole background by analyzing the properties of $V_{\text{eff}}$. The radial turning points can be found by solving the equation $\fc{dV_{\text{eff}}}{dr}=0$, and the stability of circular orbits can be investigated by examining the behavior of $V_{\text{eff}}$ near its extrema. The effective potential approach provides a clear and intuitive way to understand the dynamics of spinning test particles in the presence of a central field, such as that generated by a black hole.

\begin{figure}
	\subfigure[~$\bar{s}=0.3, \bar{j}=4$]{\label{figvr1s03j4}
		\includegraphics[width=0.22\textwidth]{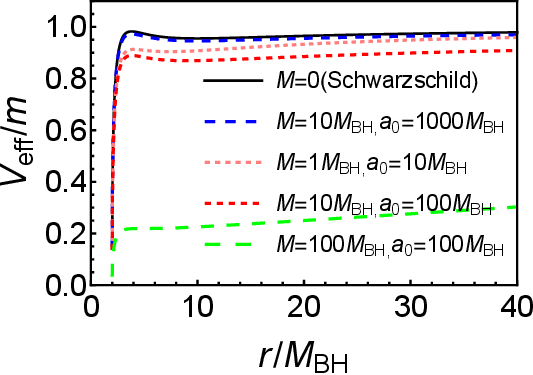}}
	\subfigure[~$\bar{j}=4, M=10M_{\text{BH}}, a_{0}=100M_{\text{BH}}$]{\label{figvr2sj4M10a100}
		\includegraphics[width=0.22\textwidth]{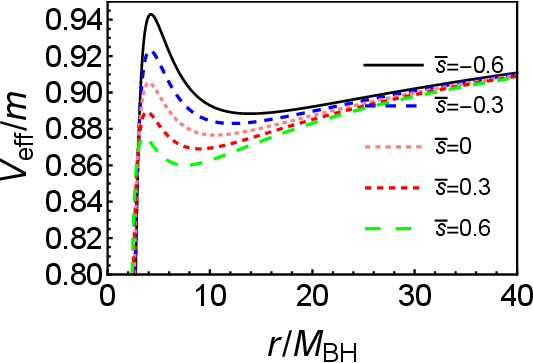}}
	\subfigure[~$ \bar{s}=0.3, M=10M_{\text{BH}}, a_{0}=100M_{\text{BH}}$]{\label{figvr3s03jM10a100}
		\includegraphics[width=0.22\textwidth]{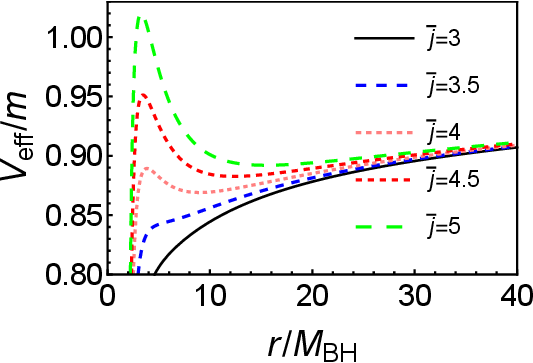}}
	\subfigure[~$\bar{s}=0.3, \bar{j}=3$]{\label{figvr4s03jM100a100}
		\includegraphics[width=0.22\textwidth]{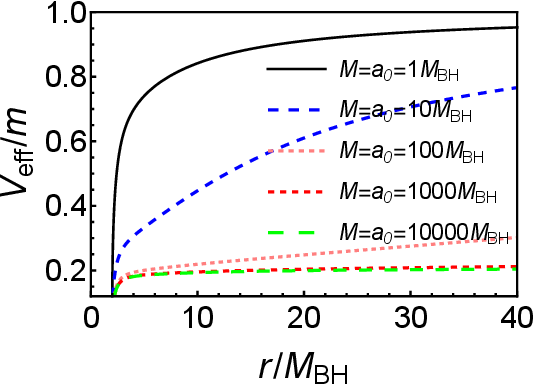}}
	\caption{The shapes of the effective potential for different parameters.}\label{figeffectivepotential1}
\end{figure}

Since the effective potential is very complicated, we only show its abstract expression. The above effective potential~\eqref{Effective_potential} with different parameters is shown in Fig.~\ref{figeffectivepotential1}. By observing the effective potential we can see how the motion of spinning test particles is affected by the dark matter halo. As can be seen from Fig.~\ref{figeffectivepotential1}, there are three cases of effective potential under different parameters: no extreme value, only a local maximum value, and both local maximum value and local minimum value. This corresponds respectively to the absence of a circular orbit, the existence of an unstable circular orbit, and the existence of both an unstable circular orbit and a stable circular orbit, respectively. Moreover, from Fig.~\ref{figvr1s03j4}, we can see that black holes with dark matter halos have lower effective potential than Schwarzschild black holes. The value of the effective potential decreases with the compactness $\mathcal{C}=\frac{M}{a_{0}}$ of the dark matter halo. Thus, spinning test particles of the same initial state follow different trajectories in the two types of black holes, and the differences widen over time. Plots of the effect of the spin angular momentum $s$ and the total angular momentum $j$ on the effective potential are shown in Figs.~\ref{figvr2sj4M10a100} and \ref{figvr3s03jM10a100}. It can be seen that the height of the effective potential decreases with the spin angular momentum $s$ and increases with the total angular momentum $j$. In addition, Fig.~\ref{figvr4s03jM100a100} shows the difference of the effective potentials at the same compactness $\mathcal{C}=1$ but with different dark matter halo masses. We can see that, although the compactness of dark matter halos is the same, the effective potential with different dark matter halos masses is still different. The larger the mass of the dark matter halo, the smaller the effective potential, that is, the more deviated from the case of the Schwarzschild black hole.

\begin{figure*}
	\subfigure[~$V_{\text{eff}}$]{\label{figvrorbit1}
		\includegraphics[width=0.3\textwidth]{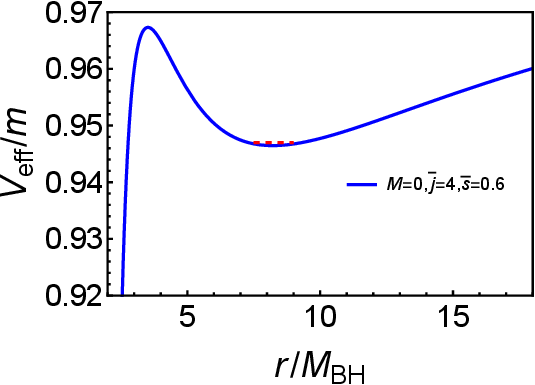}}
	\subfigure[~$V_{\text{eff}}$]{\label{figvrorbit2}
		\includegraphics[width=0.3\textwidth]{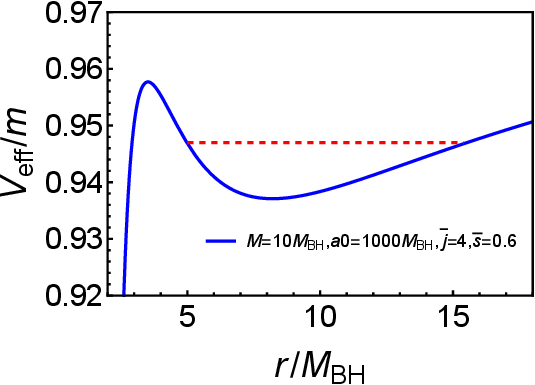}}
	\subfigure[~$V_{\text{eff}}$]{\label{figvrorbit3}
		\includegraphics[width=0.3\textwidth]{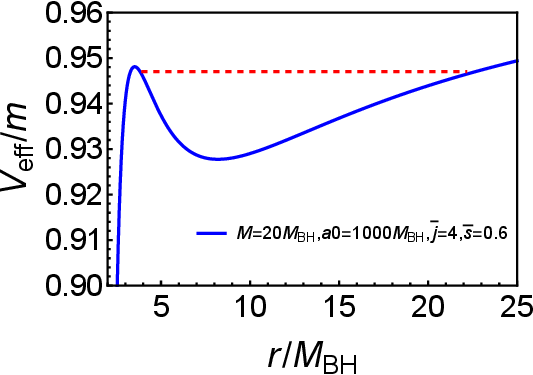}}
	\subfigure[~orbit]{\label{figObrit1}
		\includegraphics[width=0.3\textwidth]{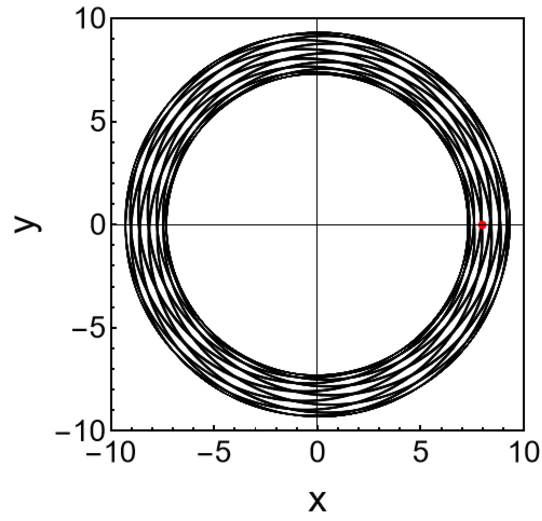}}
	\subfigure[~orbit]{\label{figObrit2}
		\includegraphics[width=0.3\textwidth]{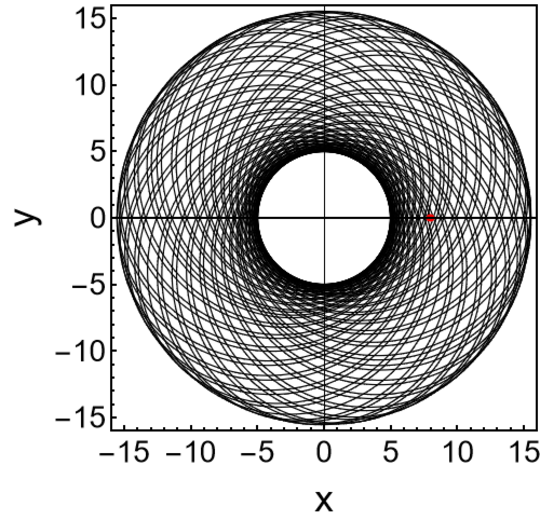}}
	\subfigure[~orbit]{\label{figObrit3}
		\includegraphics[width=0.3\textwidth]{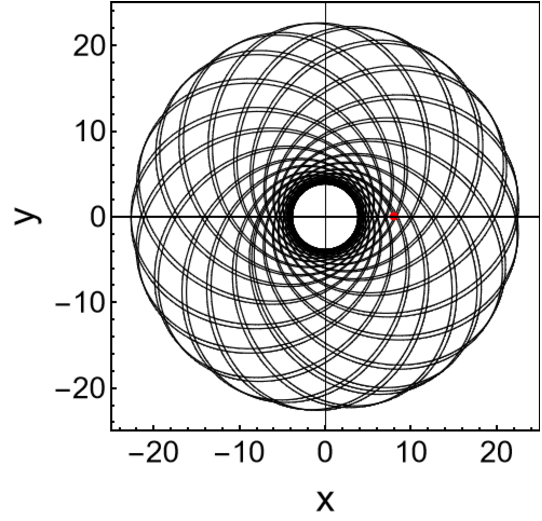}}
	\caption{Plots of the effective potential $V_{\text{eff}}$ and the corresponding orbits for a spinning test particle. The red dashed lines in the effective potentials represent that the energy of particles are $\bar{e}=0.947$. The red dots in the orbits indicate the initial position of the particles.}\label{figorbit1}
\end{figure*}

In order to more directly show the influence of dark matter halo on the motion of spinning test particles, we calculated the motion orbit of spinning test particles near the black hole with different compactness of dark matter halo. The results are shown in Fig.~\ref{figorbit1} and  Tab.~\ref{tab1}. As can be seen from Figs.~\ref{figObrit1}-\ref{figObrit3}, for a classical test particle trapped in a potential well with energy $e/m=0.947$, the radial motion range of the particle widens with the increase of the compactness of the dark matter halo. That is to say, the apastron $r_a$ increases with $\mathcal{C}$ while the periastron $r_p$ decreases with $\mathcal{C}$. We can describe the shape of the orbit by the orbital eccentricity of the orbit $\epsilon=\frac{r_a-r_p}{r_a+r_p}$. It can be seen from Tab.~\ref{tab1} that eccentricity $\epsilon$ increases with the compactness $\mathcal{C}$. Thus, the particle's motion trajectory varies significantly with the compactness of the dark matter halo.

\begin{table*}[htbp]
	\begin{tabular}{|c|c|c|c|c|}
		\hline
		$\;\;~\;\;$  &
		$\;\;r_a/M_{\text{BH}}\;\;$  &
		$\;\;\;\;\;\;\;\;r_p/M_{\text{BH}}\;\;\;\;\;\;\;$ &
		$\;\;\;\;\;\;\;\;\epsilon\;\;\;\;\;\;\;$ &
		$\;\;\;\;\;\;\;\;N\;\;\;\;\;\;\;$ \\
		\hline
		GR   &9.322~~ &7.290       &~~0.122~~& 181       \\
		$M=10M_{\text{BH}},a_0=1000M_{\text{BH}}$  &15.5382~~ &4.983       &~~0.514~~& 147       \\
		$M=20M_{\text{BH}},a_0=1000M_{\text{BH}}$   &22.646~~ &3.283       &~~0.747~~& 104       \\
		\hline
	\end{tabular}
	\caption{The apastron $r_{a}$, periastron $r_{p}$, eccentricity $\epsilon$, and number of rotations $N$ associate to the orbits in Fig.~\ref{figorbit1}.\label{tab1}}
\end{table*}

\begin{figure}
	\subfigure[~radial velocity $\dot{r}$]{\label{figusr}
		\includegraphics[width=0.22\textwidth]{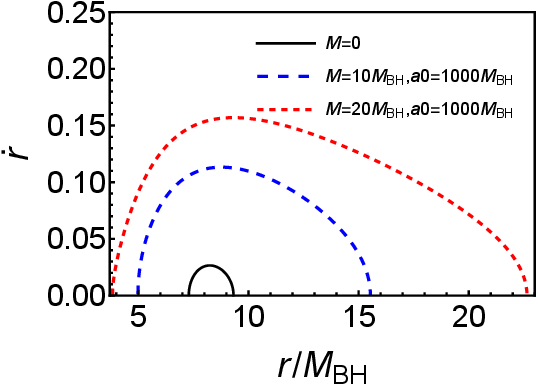}}
	\subfigure[~angular velocity $\dot{\phi}$]{\label{figusp}
		\includegraphics[width=0.22\textwidth]{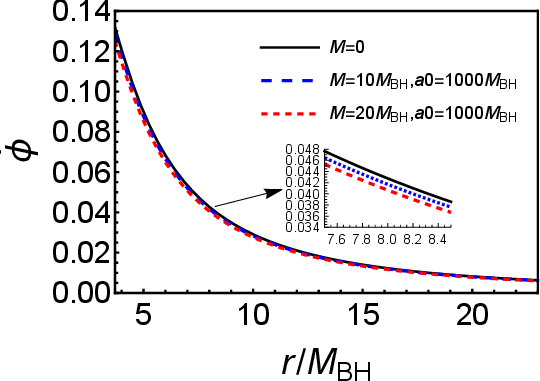}}
	\caption{Plots of the radial velocity $\dot{r}$ and angular velocity $\dot{\phi}$ for different compactness $\mathcal{C}$.}\label{figfourv}
\end{figure}

In addition, we plot the radial and angular velocities of spinning test particles at different dark matter halo densities in Fig.~\ref{figfourv}. It can be seen that radial velocity $\dot{r}$ increases with the compactness of the dark matter halo, while the angular velocity $\dot{\phi}$ decreases with the parameter $\mathcal{C}$. An interesting phenomenon is observed here. In Fig.~\ref{figusp}, we can see that the angular velocity of the spinning test particles changes little with the compactness of the dark matter halo, but Tab.~\ref{tab1} shows that the number of turns of the spinning test particles varies greatly under different compactness of the dark matter halo in the same amount of time. This is due to the fact that when the dark matter halo is more compact, the particles rotate farther in orbit on average. It can be seen from Fig.~\ref{figusr} that the angular velocity decreases with the radius $r$.

\begin{figure}
	\subfigure[~$M=0$]{\label{Causality1}
		\includegraphics[width=0.22\textwidth]{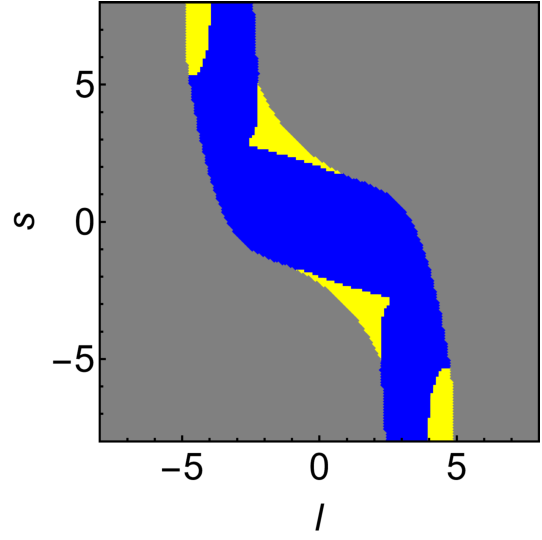}}
	\subfigure[~$M=a_{0}=100M_{\text{BH}}$]{\label{Causality2}
		\includegraphics[width=0.22\textwidth]{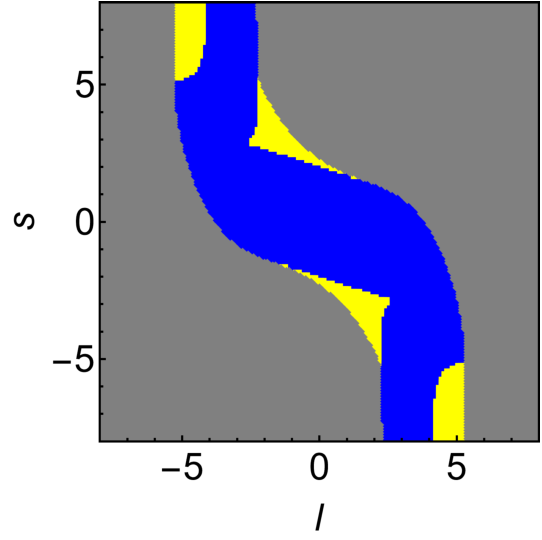}}
	\caption{Properties of the circular orbits for the spinning test particle in the $(s-l)$ parameter space. The range of $\bar{s}$ and $\bar{l}$ is $(-8, 8)$. In the gray region, the test particle can have a stable timelike circular orbit. In the yellow region, the test particle has only the unphysical spacelike circular orbits. In the blue region, the particle does not have circular orbits.}\label{Causality}
\end{figure}

At the end of this subsection, we focus on the circular orbits of spinning test particles. Since the four-velocity $u^{\mu}$ of a spin particle is not necessarily parallel to the four-momentum $P^{\mu}$, it is possible that the worldline of the spinning test particle is spacelike, which is obviously not physical. Therefore, we use superluminal constraint~\eqref{Superluminal_Constraint} to check whether the circular orbital motion of spinning test particles is superluminal. By numerical scanning, the orbital classification of black holes with different compactness in parameter space $(s-l)$ is shown in Fig.~\ref{Causality}, where $l=j-s$ is the orbit angular momentum. Similarly, $\bar{l}=\frac{l}{mM_{\text{BH}}}$ represents the dimensionless orbit angular momentum. The gray region represents that the spin particle has a time-like stable circular orbit, the blue region represents that the particle has no stable circular orbit, and the yellow region represents that the spin particle has only a space-like stable circular orbit. It can be seen that, even in the case of extremely compactness dark matter hole with large mass $M$, the diagram of parameter space does not change much from the case of the Schwarzschild black hole. This is because although the dark matter halo can affect the value of the effective potential, it does not have much effect on the overall shape of the effective potential.

\subsection{Marginally stable orbit}

\begin{figure}
	\subfigure[~$M=0$]{\label{MBOSchw}
		\includegraphics[width=0.22\textwidth]{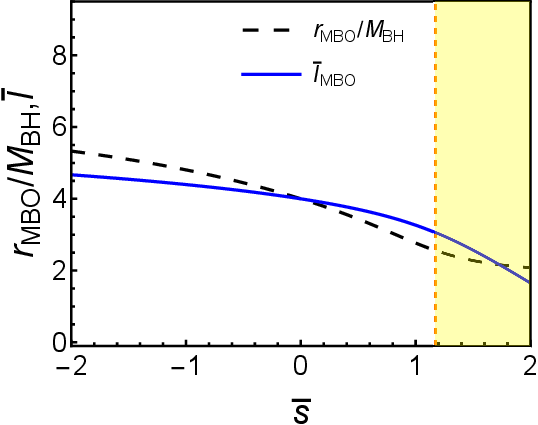}}
	\subfigure[~$M=200M_{\text{BH}}, \mathcal{C}=1/5$]{\label{MBOm200a1000}
		\includegraphics[width=0.22\textwidth]{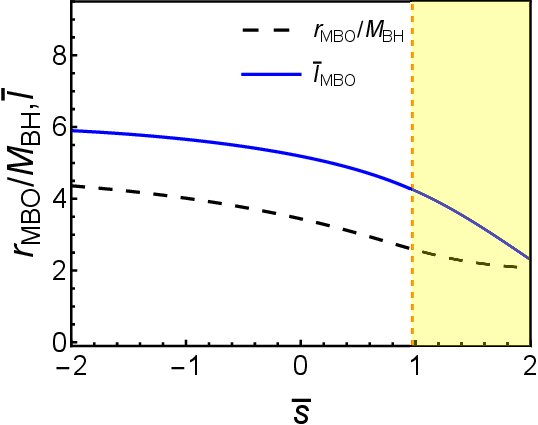}}
	\subfigure[~$\bar{s}=-0.5$]{\label{MBOsn05a1000}
	\includegraphics[width=0.22\textwidth]{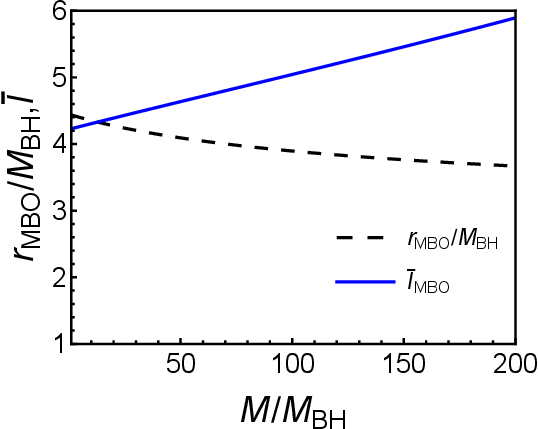}}
\subfigure[~$\bar{s}=0.5$]{\label{MBOs05a1000}
	\includegraphics[width=0.22\textwidth]{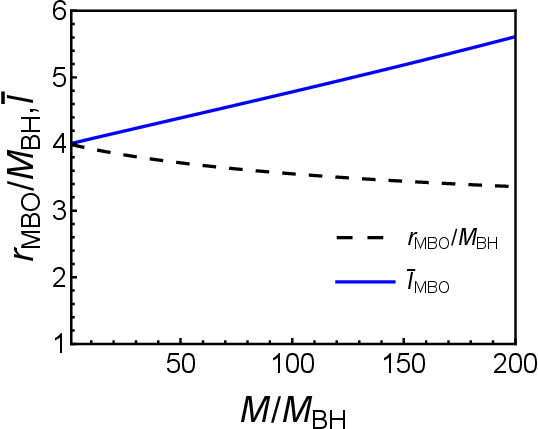}}		
	\caption{Plots of the radius $r_{\text{MBO}}$ and orbital angular momentum $l_{\text{MBO}}$ of the spinning test particle for the different parameters $\mathcal{C}$ and $s$. The light yellow areas indicate that the particle orbits are spacelike.}\label{figmbo}
\end{figure}

For a spinning particle, the MBO represents the smallest possible circular bound orbit with the smallest radius and an energy of $\bar{e} = 1$. Given an effective potential $V_{\rm eff}$ controlling spinning particle motion, MBO satisfies the conditions~\cite{Wei:2019zdf} 
\begin{eqnarray}
V_{\rm eff} = 1  \\
 dV_{\rm eff}/dr = 0.
\end{eqnarray}
The radius and orbital angular momentum of MBO can be obtained by solving these equations. The radius $r_{\rm MBO}$ and angular momentum $l_{\rm MBO}$ are obviously dependent on the dark matter halo compactness $\mathcal{C}$ and the particle spin $s$. Figure~\ref{figmbo} shows the numerical results of $r_{\rm MBO}$ and $l_{\rm MBO}$ as functions of the particle spin parameter $s$ and the dark matter mass $M$. We observe that both $r_{\rm MBO}$ and $l_{\rm MBO}$ decrease with $s$. When $\mathcal{C}$ increases, $r_{\rm MBO}$ decreases and $l_{\rm MBO}$ increases.

\subsection{ISCO}

\begin{figure}
	\subfigure[~$M=0$]{\label{figrvmaxrvminsch}
		\includegraphics[width=0.22\textwidth]{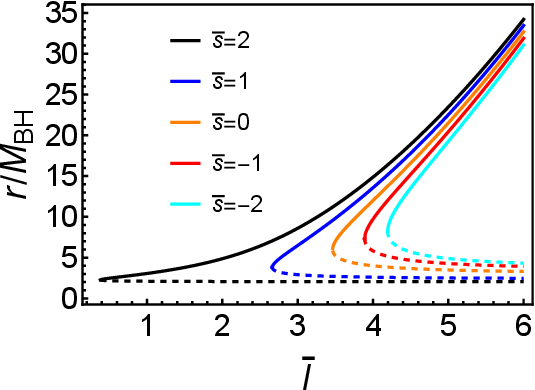}}
	\subfigure[~$M=a_{0}=100M_{\text{BH}}$]{\label{figrvmaxrvminm100a100}
		\includegraphics[width=0.22\textwidth]{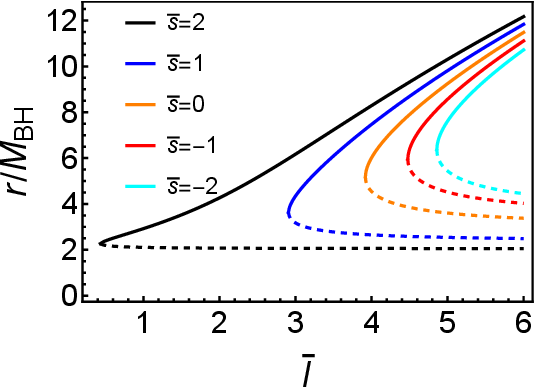}}
	\caption{The relation between the position of the minimum and maximum of the effective potential and the orbital angular momentum $l$. The solid line represents the minimum value (stable circular orbit) and the dashed line represents the maximum value (stable circular orbit).}\label{figrvmaxrvmin}
\end{figure}

Now, we consider the ISCO of a spinning test particle around a black hole in a dark matter halo. The ISCO marks the inner edge of the  accretion disk of a black hole, so it is important from an observational perspective. A particle with a circular orbit satisfies the following conditions: 
\begin{eqnarray}
\fc{dr}{d\lambda}&=&0,\label{iscocondition1}\\
\fc{d^2r}{d\lambda^2}&=&0~~~~\text{or}~~~~\fc{dV_{\text{eff}}}{dr}=0,\label{iscocondition2}
\end{eqnarray}
the above conditions correspond to the particle's radial velocity and radial acceleration vanish, respectively. But that's just for a circular orbit. If it's a stable circular orbit, it needs to satisfy another condition: $\fc{d^2V_{\text{eff}}}{dr^2}\leq0$. The ISCO of a spin particle is the stable circular orbit with the smallest radius, and thus corresponds to the position where the maximum and minimum of the effective potential of the particle become the same point, i.e
\begin{equation}
\fc{d^2V_{\text{eff}}}{dr^2}=0.\label{iscocondition3}
\end{equation}
In Fig.~\ref{figrvmaxrvmin}, we show the relation between the radius $r$ of the circular orbit and the angular momentum $l$ with different spins $s$. The solid line represents the radius of the unstable circular orbit and the dashed line represents the radius of the stable circular orbit.  The intersection of the solid line and the dashed line is the ISCO of the spin particle. Obviously, the radius of the ISCO decreases with the spin angular momentum $s$ of the particle.

\begin{figure}
	\subfigure[~$M=0$]{\label{ISCOSchw}
		\includegraphics[width=0.22\textwidth]{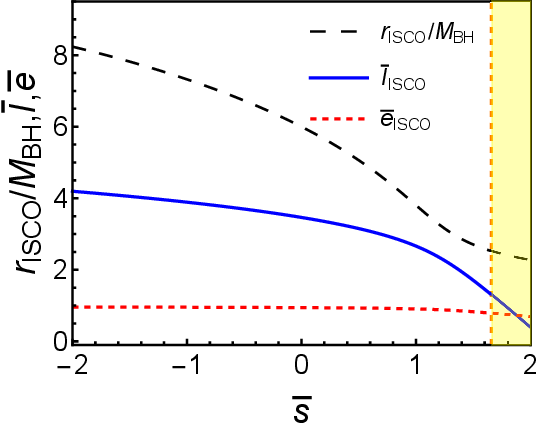}}
	\subfigure[~$M=10M_{\text{BH}}, \mathcal{C}=1/100$]{\label{ISCOm10a1000}
		\includegraphics[width=0.22\textwidth]{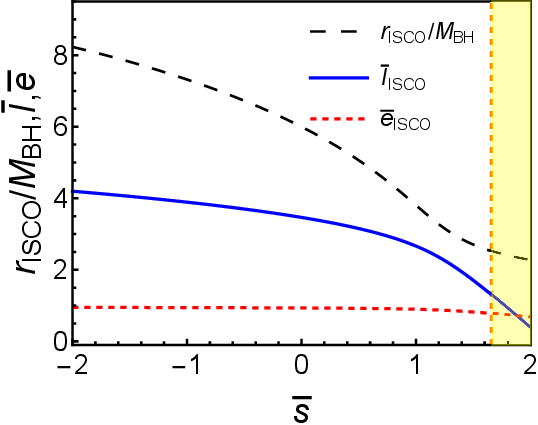}}
	\subfigure[~$M=100M_{\text{BH}}, \mathcal{C}=1/10$]{\label{ISCOm100a1000}
		\includegraphics[width=0.22\textwidth]{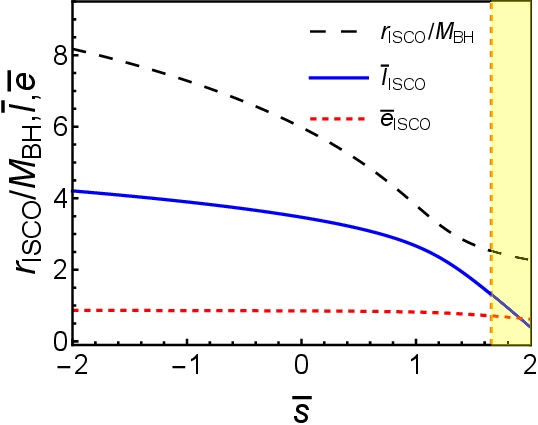}}
	\subfigure[~$M=100M_{\text{BH}}, \mathcal{C}=1$]{\label{ISCOm100a100}
		\includegraphics[width=0.22\textwidth]{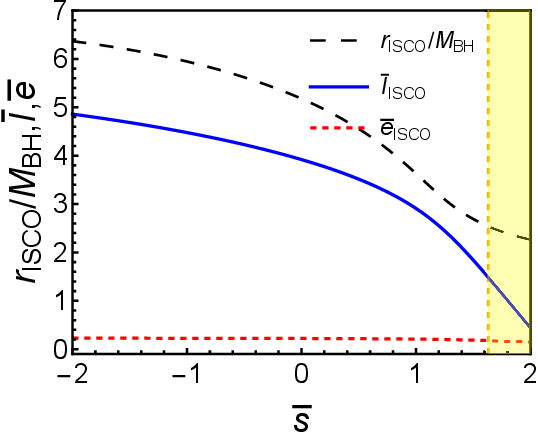}}
	\caption{Plots of the radius $r_{\text{ISCO}}$, orbital angular momentum $l_{\text{ISCO}}$, and energy $e_{\text{ISCO}}$ of the spinning test particle for the different parameters $\mathcal{C}$ and $s$. The light yellow areas indicate that the particle orbits are spacelike.}\label{figISCOma}
\end{figure}

\begin{figure}
	\subfigure[~$\bar{s}=-1$]{\label{ISCOsn1a100}
		\includegraphics[width=0.22\textwidth]{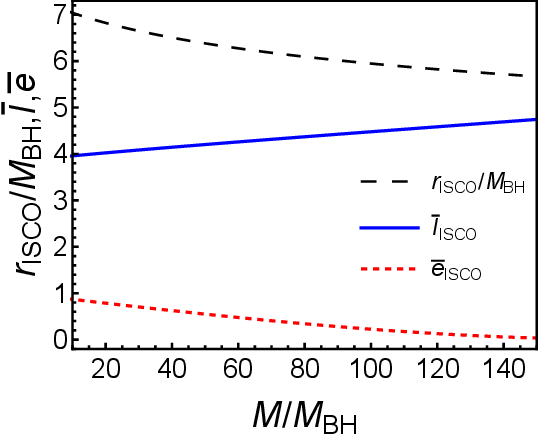}}
			\subfigure[~$\bar{s}=-0.5$]{\label{ISCOsn05a100}
			\includegraphics[width=0.22\textwidth]{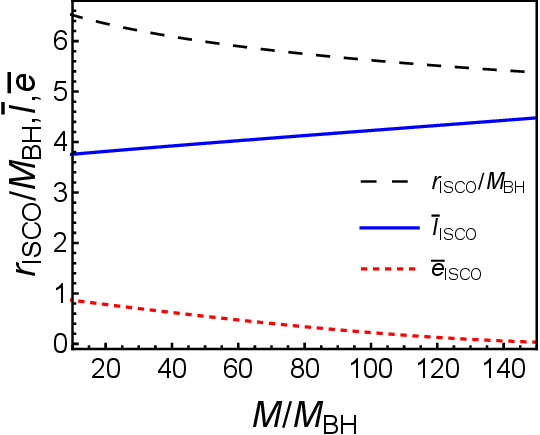}}
	\subfigure[~$\bar{s}=0.5$]{\label{ISCOs05a100}
		\includegraphics[width=0.22\textwidth]{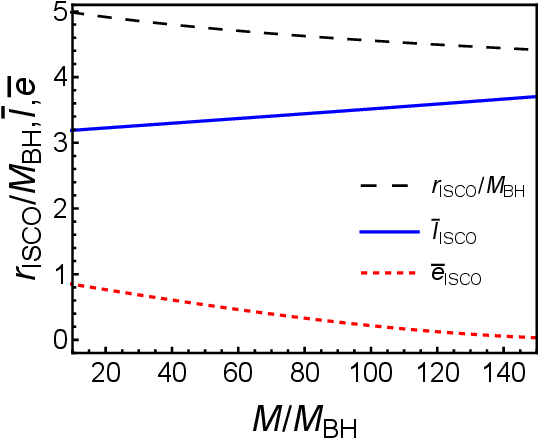}}
	\subfigure[~$\bar{s}=1$]{\label{ISCOs1a100}
		\includegraphics[width=0.22\textwidth]{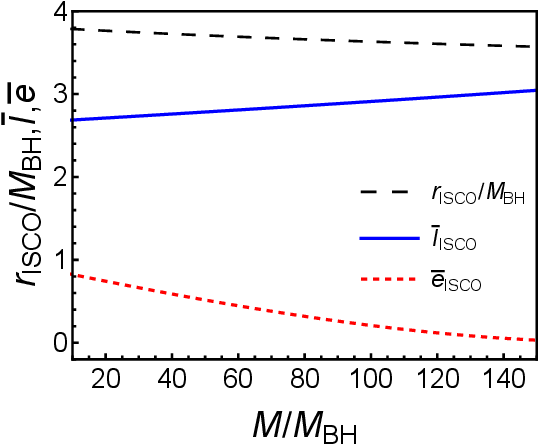}}
	\caption{Plots of the radius $r_{\text{ISCO}}$, orbital angular momentum $l_{\text{ISCO}}$, and energy $e_{\text{ISCO}}$ of the spinning test particle for the different parameters $s$ and $M$. The parameter $a_{0}$ is set to $a_{0}=100M_{\text{BH}}$.}\label{figISCOsa}
\end{figure}

\begin{figure}
	\subfigure[~$\bar{s}=-1$]{\label{ISCOs0M10}
		\includegraphics[width=0.22\textwidth]{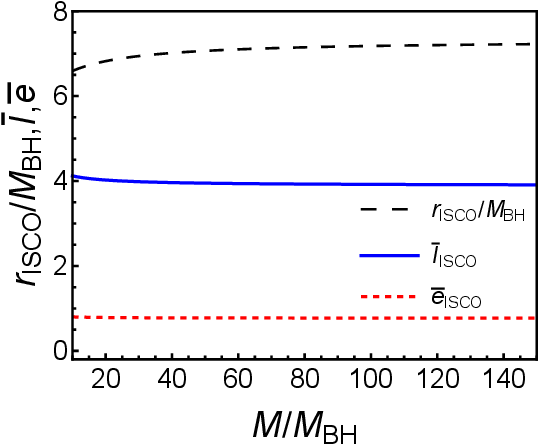}}
	\subfigure[~$\bar{s}=-0.5$]{\label{ISCOs05M10}
		\includegraphics[width=0.22\textwidth]{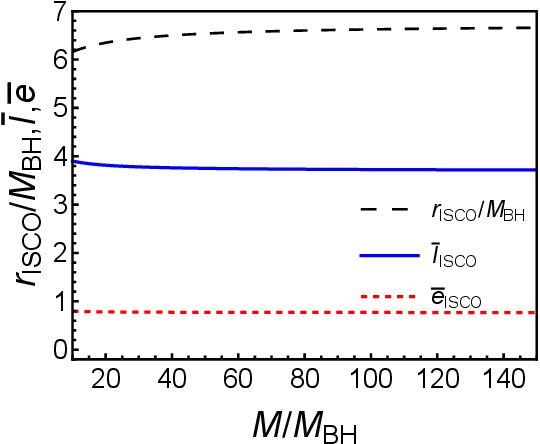}}
	\subfigure[~$\bar{s}=0.5$]{\label{ISCOs1M10}
		\includegraphics[width=0.22\textwidth]{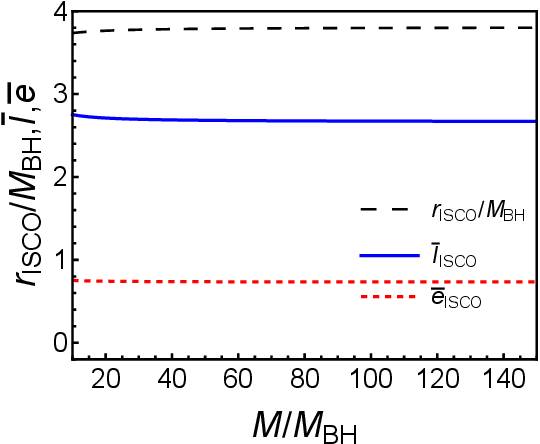}}
	\subfigure[~$\bar{s}=1$]{\label{ISCOs15M10}
		\includegraphics[width=0.22\textwidth]{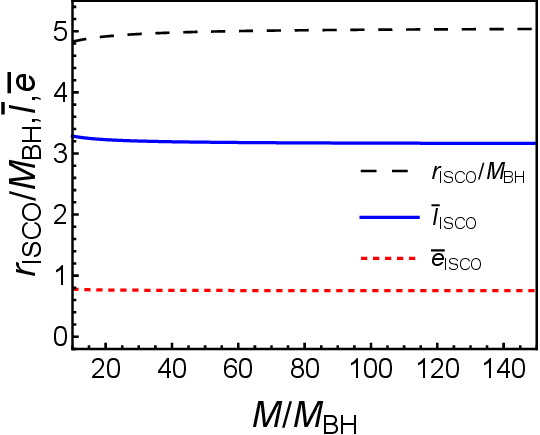}}
	\caption{Plots of the radius $r_{\text{ISCO}}$, orbital angular momentum $l_{\text{ISCO}}$, and energy $e_{\text{ISCO}}$ of the spinning test particle for the  parameters $M$ and $s$. The parameter $\mathcal{C}$ is set to $\mathcal{C}=0.2$. }\label{figISCOsm}
\end{figure}

Next, we use the above three conditions \eqref{iscocondition1}, \eqref{iscocondition2}, \eqref{iscocondition3}, and the superluminal constraint~\eqref{Superluminal_Constraint} to solve the ISCO of the spinning test particle. In Figs.~\ref{figISCOma}, ~\ref{figISCOsm}, and ~\ref{figISCOsa}, we plot the radius $r_\text{ISCO}$, orbital angular momentum $\bar{l}_\text{ISCO}$, and energy $\bar{e}_\text{ISCO}$ with different parameters. From Fig.~\ref{figISCOma}, we can see that the ISCO radius $r_{\text{ISCO}}$, orbital angular momentum $\bar{l}_\text{ISCO}$, and energy $\bar{e}_\text{ISCO}$ of spinning test particles all decrease with the spin $s$. When the spin $s$ is large, the trajectory of particle becomes spacelike, as shown in the yellow region in Fig.~\ref{figISCOma}.

In addition to the particle spin $s$, the parameters of the background black hole also affect the ISCO of the particle. In Fig.~\ref{figISCOsa}, we show the variation of the parameters of the ISCO with the mass $M$ of the dark matter halo when the radius of dark matter halo is fixed. It can be seen that, the radius $r_{\text{ISCO}}$ and the particle energy $s$ decrease with $M$, and the orbital angular momentum $j$ of the particle increases with $M$. This is equivalent to the fact that the greater the compactness $\mathcal{C}$ of the dark matter halo, the smaller the radius $r_{\text{ISCO}}$ and the particle energy $s$. However, as Fig.~\ref{figvr4s03jM100a100} points out, although the dark matter halo has the same compactness, it could still has different masses, and the effective potential corresponding to the black hole is also different. Therefore, we also study the ISCO of spinning test particles of black holes with same compactness $\mathcal{C}$ for different masses of dark matter halos, and the results are shown in Fig.~\ref{figISCOsm}. It can be seen that although the compactness of the dark matter halo is the same, the ISCO parameters are still slightly different under different dark matter halo masses. This shows that compactness alone cannot fully reflect the properties of orbital motion of spinning test particles in black holes with dark matter halos. The above results show that, dark matter halos affect the orbital motion of spinning test particles. The inner edge of the accretion disk of a black hole with a dark matter halo is smaller and closer to the black hole's event horizon than a Schwarzschild black hole. We expect future observations to use this result to probe or limit the density of dark matter halos.

\subsection{periodic orbits}

After investigating the ISCOs and MBOs, we now shift our focus to periodic orbits near a black hole surrounded by dark matter halos. These orbits are characterized by their repetitive trajectories, which arise when the frequency ratio between the radial ($\omega_r$) and azimuthal ($\omega_\phi$) oscillations is rational. Although generic orbits may exhibit irrational frequency ratios, they can always be approximated by nearby periodic orbits with arbitrary precision.
	
\begin{figure}
	\subfigure[~$M=0$]{\label{qESchc}
		\includegraphics[width=0.22\textwidth]{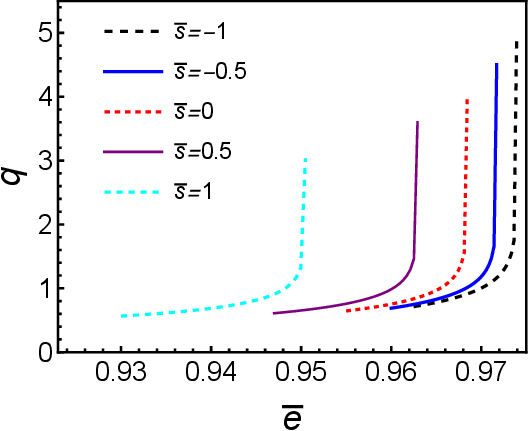}}
	\subfigure[~$\bar{s}=0.5, a_{0}=1000M_{\text{BH}}$]{\label{qEM}
		\includegraphics[width=0.22\textwidth]{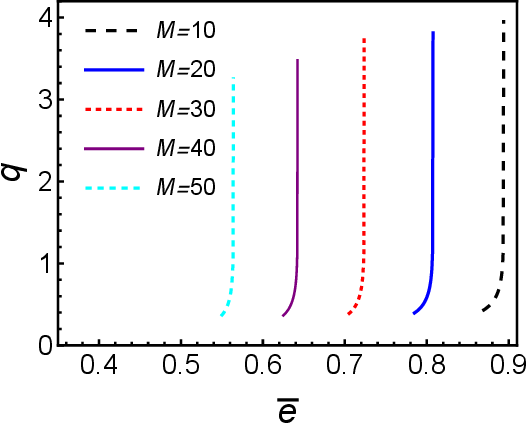}}
	\caption{Plots of the relations of the rational number $q$ and the energy $\bar{e}$ for different parameters $s$ and $\mathcal{C}$. The orbit angular momentum $l$ set to $l=\frac{l_{\text{MBO}}+l_{\text{ISCO}}}{2}$.}\label{qEplot}
\end{figure}

In the spherically symmetric geometry under consideration, the motion is governed by radial $r$-motion and angular $\phi$-motion. The apsidal angle, $\Delta\phi$, is calculated by integrating the change in azimuthal angle over one radial oscillation, from periapsis ($r_1$) to apoapsis ($r_2$) and back~\cite{Levin:2008mq}
	\begin{eqnarray}
		\Delta\phi=\oint d\phi &=&2\int_{\phi_1}^{\phi_2} d\phi=2 \int_{r_1}^{r_2} \frac{\dot{\phi}}{\dot{r}}dr.
\end{eqnarray}
The factor of 2 accounts for the symmetrical path followed by the particle. The apsidal angle is significantly influenced by the particle's energy, angular momentum, and the spacetime structure around the black hole, as indicated by the presence of functions $a(r)$ and $b(r)$. Consequently, black holes with different parameter values will exhibit varying apsidal angles. For the bound orbital we're studying, the angular momentum $l$ is only between $l_{\text{ISCO}}$ and $l_{\text{MBO}}$, which is
\begin{equation}
l_{\text{ISCO}}<l<l_{\text{MBO}},
	\end{equation}
Once the orbital angular momentum and the energy of the particle are given, we can find the periodic orbit of the spin particle by the frequency ratio $q$. The frequency ratio $q$ is defined as the relation between $\omega_r$ and $\omega_\phi$, expressed in terms of three integers: zoom $z$, whirl $w$, and vertex $v$~\cite{Levin:2008mq}
	\begin{equation}\label{qradial}
		q = \frac{\omega_\phi}{\omega_r} - 1 =\frac{\Delta\phi}{2\pi}-1= w + \frac{v}{z}.
	\end{equation}
For irrational values of $q$, the particle's trajectory traces a precessing orbit, characterized by a precession angle $w = \Delta\Phi - 2\pi$. Conversely, rational values of $q$ correspond to periodic orbits, where the particle returns to its initial position after a finite time. To illustrate the behavior of $q$ more directly, we plot the relation between $q$ and the energy $e$ of spinning test particle for different parameters in Fig.~\ref{qEplot}. It can be seen that, the rational number $q$ increases with the $\bar{e}$. In Fig.~\ref{qESchc}, we take the black hole as Schwarzschild case but take different values of spin angular momentum $s$, and it can be seen that $s$ has a significant effect on the distribution of $q-e$. In Fig.~\ref{qEM}, we fixed $\bar{s}=0.5$ and explored the effect of different dark matter halo masses $M$ on $q$. The results show that the closer the system is to the Schwarzschild case, the greater the maximum energy corresponding to $q$.
	
\begin{figure*}
	\subfigure[~$\bar{e}=0.959889$]{\label{orbit110}
		\includegraphics[width=0.22\textwidth]{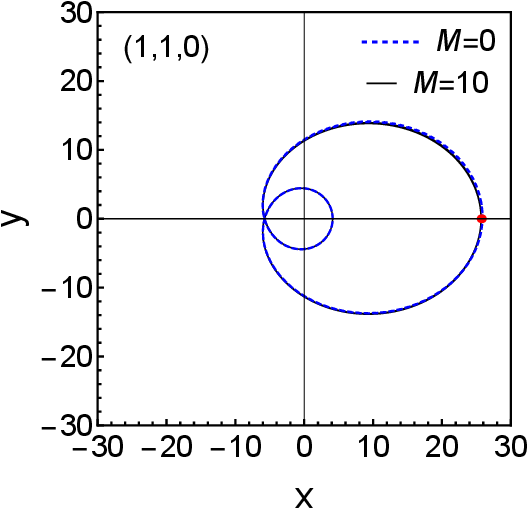}}
	\subfigure[~$\bar{e}=0.962253$]{\label{orbit120}
		\includegraphics[width=0.22\textwidth]{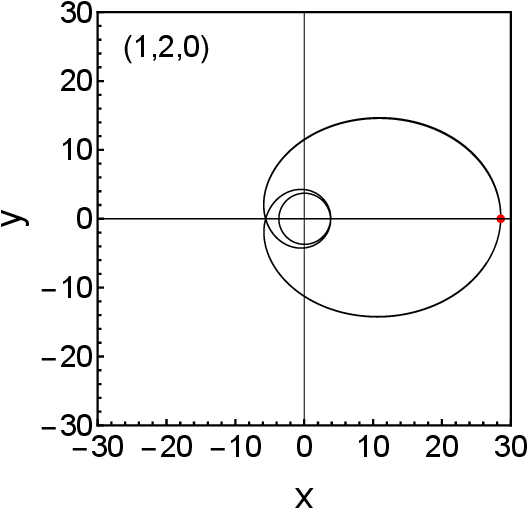}}
			\subfigure[~$\bar{e}=0.962015$]{\label{orbit211}
			\includegraphics[width=0.22\textwidth]{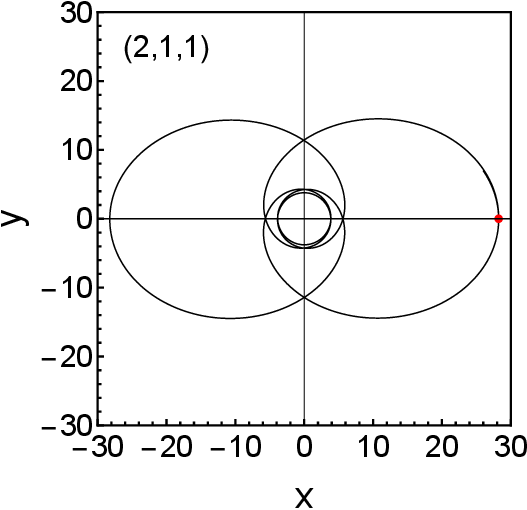}}
				\subfigure[~$\bar{e}=0.962281$]{\label{orbit221}
				\includegraphics[width=0.22\textwidth]{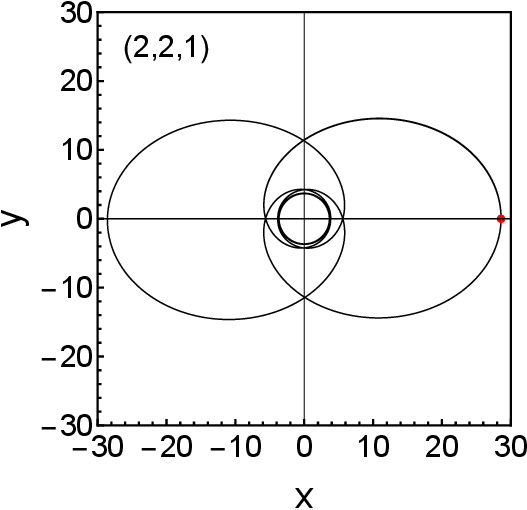}}
					\subfigure[~$\bar{e}=0.962152$]{\label{orbit312}
					\includegraphics[width=0.22\textwidth]{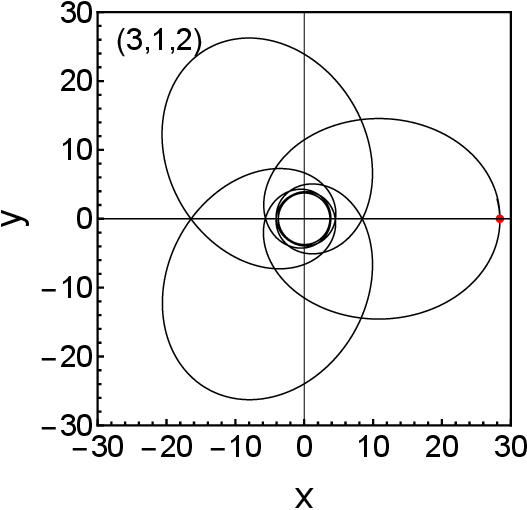}}
				\subfigure[~$\bar{e}=0.962283$]{\label{orbit322}
					\includegraphics[width=0.22\textwidth]{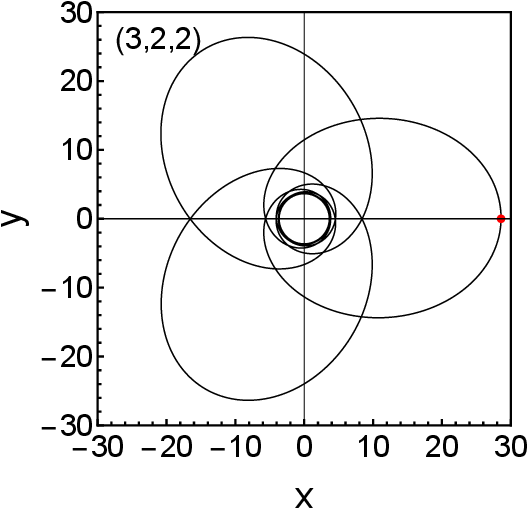}}
				\subfigure[~$\bar{e}=0.954769$]{\label{orbit403}
					\includegraphics[width=0.22\textwidth]{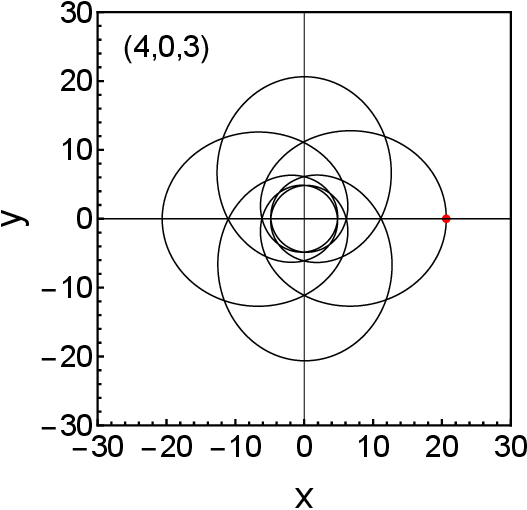}}
				\subfigure[~$\bar{e}=0.962192$]{\label{orbit413}
					\includegraphics[width=0.22\textwidth]{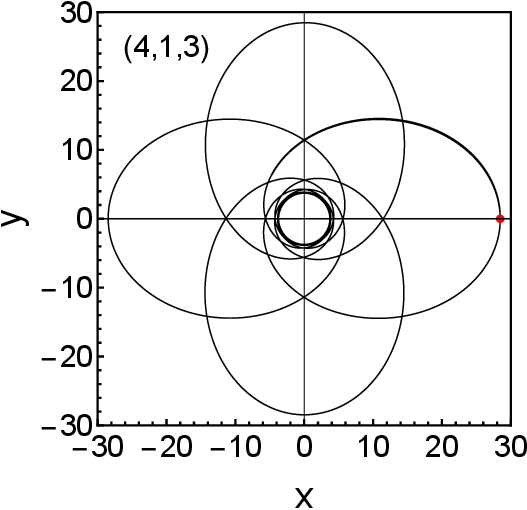}}
	\caption{Periodic orbits for different values of $(z, w, v)$ around a black hole with dark matter halo. The value of parameters are $M=10M_{BH}$, $a0 = 1000M_{BH}$, $\bar{s}=0.5$, and $\bar{l}=4$.}\label{periodic orbit}
\end{figure*}

Now we study the orbits of the spin particles in specific periodic orbits. Figure~\ref{periodic orbit} shows periodic orbits using different combinations of integers $(z, w, v)$, with the value of $z$ determining the number of blades in the orbit's shape. Larger $z$ values correspond to more complex trajectories with larger blade profiles. From Fig.~\ref{orbit110}, we can see that when the dark matter halo parameters are $M=10$ and $a_0=1000$, the periodic orbits of the spin particles almost coincide with those in the Schwarzschild black hole case. But we believe that as the compactness $\mathcal{C}$ of the dark matter halo increases, the difference becomes apparent. Moreover, even if the periodic orbits of the spin particles around the dark matter halo black hole are very similar to those of the Schwarzschild black hole, because of the long time scale of the orbit and the radiation of gravitational waves, this may cause the initial small difference to become apparent later. Therefore, there is a good chance that future highly sensitive gravitational wave detectors or telescopes will detect this difference, which will further validate or constrain black hole models with dark matter halos.

\section{Conclusions}\label{sec:conclusions}

In this paper, we investigated the equatorial plane motion of a spinning test particle around the black hole immersed in a dark matter halo. Because of the spin-curvature coupling, the equation of particle motion is described by the MPD equation rather than the geodesic equation. Considering the simple case of parallel or anti-parallel spin momentum $s$ and orbital angular momentum $j$, we studied how dark matter halos affect the motion of spinning test particles using the MPD equations. We mainly studied three kinds of orbits, namely MBO, ISCO, and periodic orbit. We found that the dark matter halo can make the MBO and ISCO of the spin particle smaller, that is, the inner edge of the accretion disk of the black hole with the dark matter halo closer to the event horizon. This is consistent with the conclusion of the ISCO of massless particles around black holes with dark matter halos~\cite{Cardoso:2021wlq}. Our research contributes to a deeper understanding of the complex gravitational interactions between black holes and dark matter halos, providing promising possibilities for future observations of the effects of dark matter in strong gravitational fields.

We began by reviewing the solution for a spherically symmetric black hole surrounded by dark matter halos. Based on this solution, we use the MPD equation and Tulczyjew spin-supplementary condition to obtain the effective potential and four-velocity of the spinning test particle moving along the equatorial plane. By using the effective potential and four velocities, we obtain the orbital characteristics of the spinning test particles, which is shown in Fig.~\ref{figorbit1} and Tab.~\ref{tab1}. The results showed that the apastron $r_a$ and eccentricity $\epsilon$ increase with $\mathcal{C}$ while the periastron $r_p$ and the number of turns $N$ decrease with $\mathcal{C}$. This suggests that dark matter halos do affect the orbits of spinning test particles. In addition, Fig.~\ref{Causality} shows a parametric diagram indicating whether a spinning particle has a circular orbit in the $s-l$ parameter space. We found that compared with the Schwarzschild black hole, the difference in the parameter graph of black holes with dark matter halos is not obvious. This is due to the small influence of dark matter halos on the overall shape of the effective potential. We studied the MBO, ISCO, and periodic orbit of the spinning test particles. From Fig.~\ref{figmbo}, we found that the spin $s$ of a particle and the mass of dark matter halos significantly affect the radius $r_{\text{MBO}}$, orbital angular momentum $l_{\text{MBO}}$ of the spinning particle. The same effect is observed for the ISCO. However, when the spin of the particle is large, the trajectories of the particle may become spacelike. By investigating the influence of the compactness $\mathcal{C}$ on the ISCO of the black hole, we found that dark matter halos make the $r_{\text{ISCO}}$ of a spinning test particle smaller, and the more compact the dark matter halo, the smaller the radius. And dark matter halos of the same compactness but different masses have slightly different results, which is shown in Figs.~\ref{figISCOsa} and~\ref{figISCOsm}. Finally, the study of periodic orbits of spin particles shows that spin particles have periodic orbital motions just like non-spin particles. The presence of dark matter would change the orbital characteristics of periodic orbits. These results showed that dark matter halos do have an effect on the motion of spinning test particles around the black hole. Although dark matter halos in the real world are less compact, increasingly accurate astronomical observations may detect these faint signs in the near future.

Our work could be improved in a number of ways. For example, we could consider the influence of the black hole's spin and explore other spin-supplementary conditions besides the Tulczyjew condition. Additionally, we could investigate the motion of spinning test particles in a black hole background with dark matter by the Jacobi elliptic functions.

\vspace{5mm}
\begin{acknowledgments}

This work was supported by the National Natural Science Foundation of China (Grants No. 12035005, No. 12405055, and No. 12347111), the China Postdoctoral Science Foundation (Grant No. 2023M741148), the Postdoctoral Fellowship Program of CPSF (Grant No. GZC20240458), and the National Key Research and Development Program of China (Grant No. 2020YFC2201400).

\end{acknowledgments}


\begin{thebibliography}{80}


\bibitem{Dietrich:2012}
J. P. Dietrich, N. Werner, D. Clowe, A. Finoguenov, T. Kitching, L. Miller, and A. Simionescu,
{\it {A filament of dark matter between two clusters of galaxies}},
{\em Nature} {\bf 487}, 202  (2012),
[{{\tt arXiv:1207.0809}}].



\bibitem{Planck:2018vyg}
N.~Aghanim \textit{et al.} [Planck],
{\it {Planck 2018 results. VI. Cosmological parameters}},
{\em Astron. Astrophys.} \textbf{641},  A6 (2020),
[erratum: Astron. Astrophys. \textbf{652}, C4  (2021)],
[{{\tt arXiv:1807.06209 }}].



\bibitem{Begeman:1991iy}
K.~G.~Begeman, A.~H.~Broeils and R.~H.~Sanders,
{\it {Extended rotation curves of spiral galaxies: Dark haloes and modified dynamics}},
{\em Mon. Not. Roy. Astron. Soc.} \textbf{249}, 523  (1991).



\bibitem{Sadeghian:2013laa}
L.~Sadeghian, F.~Ferrer, and C.~M.~Will,
{\it {Dark matter distributions around massive black holes: A general relativistic analysis}},
{\em Phys. Rev. D} \textbf{88}, 063522  (2013),
[{{\tt arXiv:1305.2619}}].




\bibitem{Barack:2018yly}
L.~Barack, V.~Cardoso, S.~Nissanke, T.~P.~Sotiriou, A.~Askar, C.~Belczynski, G.~Bertone, E.~Bon, D.~Blas, and R.~Brito, \textit{et al.}
{\it {Black holes, gravitational waves and fundamental physics: a roadmap}},
{\em Class. Quant. Grav.} \textbf{36}, 143001  (2019),
[{{\tt arXiv:1806.05195}}].




\bibitem{Bertone:2018krk}
G.~Bertone and T. M. P.~Tait,
{\it {A new era in the search for dark matter}},
{\em Nature} \textbf{562}, 51  (2018),
[{{\tt arXiv:1810.01668}}].



\bibitem{LIGOScientific:2016aoc}
B.~P.~Abbott {\it et al.} [LIGO Scientific and Virgo],
{\it {Observation of Gravitational Waves from a Binary Black Hole Merger}},
{\em Phys. Rev. Lett.} {\bf 116}, 061102 (2016),
[{{\tt arXiv:1602.03837}}].

\bibitem{LIGOScientific:2020ibl}
R.~Abbott {\it et al.} [LIGO Scientific and Virgo],
{\it {GWTC-2: Compact Binary Coalescences Observed by LIGO and Virgo During the First Half of the Third Observing Run}},
{\em Phys. Rev. X} {\bf 11}, 021053 (2021),
[{{\tt arXiv:2010.14527}}].



\bibitem{EventHorizonTelescope:2019dse}
K.~Akiyama {\it et al.} [Event Horizon Telescope],
{\it {First M87 Event Horizon Telescope Results. I. The Shadow of the Supermassive Black Hole}},
{\em Astrophys. J. Lett.} {\bf 875}, L1 (2019),
[{{\tt arXiv:1906.11238}}].



\bibitem{Cardoso:2021wlq}
V.~Cardoso, K.~Destounis, F.~Duque, R.~P.~Macedo, and A.~Maselli,
{\it {Black holes in galaxies: Environmental impact on gravitational-wave generation and propagation}},
{\em Phys. Rev. D} {\bf 105}, L061501 (2022),
[{{\tt arXiv:2109.00005}}].


\bibitem{Hernquist:1990}
L.~Hernquist,
{\it {An analytical model for spherical galaxies and bulges}},
{\em Astrophys. J.} {\bf 356}, 359 (1990).



\bibitem{Konoplya:2021ube}
R.~A.~Konoplya,
{\it {Black holes in galactic centers: Quasinormal ringing, grey-body factors and Unruh temperature}},
{\em Phys. Lett. B} {\bf 823}, 136734 (2021),
[{{\tt arXiv:2109.01640}}].


\bibitem{Xavier:2023exm}
S.~V.~M.~C.~B.~Xavier, H.~C.~D.~Lima, Junior., and L.~C.~B.~Crispino,
{\it {Shadows of black holes with dark matter halo}},
{\em Phys. Rev. D} {\bf 107}, 064040 (2023),
[{{\tt arXiv:2303.17666}}].


\bibitem{Liu:2022lrg}
J.-Y.~Liu, S.-B.~Chen, and J.-J.~Jing,
{\it {Tidal effects of a dark matter halo around a galactic black hole*}},
{\em Chin. Phys. C} {\bf 46}, 105104 (2022),
[{{\tt arXiv:2203.14039}}].

\bibitem{LimaJunior:2022gko}
H.~C.~D.~Lima, Junior, M.~M.~Corr\^ea, C.~F.~B.~Macedo, and L.~C.~B.~Crispino,
{\it {Tidal forces in dirty black hole spacetimes}},
{\em Eur. Phys. J. C} {\bf 82}, 479 (2022),
[{{\tt arXiv:2205.13569}}].



\bibitem{Cardoso:2022whc}
V.~Cardoso, K.~Destounis, F.~Duque, R.Panosso Macedo, and A.~Maselli,
{\it {Gravitational Waves from Extreme-Mass-Ratio Systems in Astrophysical Environments}},
{\em Phys. Rev. Lett.} {\bf 129}, 241103 (2022),
[{{\tt arXiv:2210.01133}}].

\bibitem{Figueiredo:2023gas}
E.~Figueiredo, A.~Maselli, and V.~Cardoso,
{\it {Black holes surrounded by generic dark matter profiles: Appearance and gravitational-wave emission}},
{\em Phys. Rev. D} {\bf 107}, 104033 (2023),
[{{\tt arXiv:2303.08183}}].










\bibitem{Jusufi:2022jxu}
K.~Jusufi,
{\it {Black holes surrounded by Einstein clusters as models of dark matter fluid}},
{\em Eur. Phys. J. C} {\bf 83}, 103 (2023),
[{{\tt arXiv:2202.00010}}].

\bibitem{Konoplya:2022hbl}
R.~A.~Konoplya and A.~Zhidenko,
{\it {Solutions of the Einstein Equations for a Black Hole Surrounded by a Galactic Halo}},
{\em Astrophys. J.} {\bf 933}, 166 (2022),
[{{\tt arXiv:2202.02205}}].



\bibitem{Shen:2024qxv}
Z.-B.~Shen, A.-Z.~Wang, and S.-Y.~Yin,
{\it {A Class of Analytical Models for Black holes Surrounded by Dark Matter Halos}},
[{{\tt arXiv:2408.05417}}].



\bibitem{Stelea:2023yqo}
C.~Stelea, M.~A.~Dariescu, and C.~Dariescu,
{\it {Charged black holes with dark halos}},
{\em Phys. Lett. B}  {\bf 847}, 138275  (2023),
[{{\tt arXiv:2309.13651}}].





\bibitem{Zhang:2024hjr}
C.~Zhang, G.-Y.~Fu, and C.-Y.~Zhang,
{\it {Rotating galactic black holes}},
[{{\tt arXiv:2403.19933}}].




\bibitem{Babak:2017tow}
S.~Babak, J.~Gair, A.~Sesana, E.~Barausse, C.~F.~Sopuerta, C.~P.~L.~Berry, E.~Berti, P.~Amaro-Seoane, A.~Petiteau, and A.~Klein,
{\it {Science with the space-based interferometer LISA. V: Extreme mass-ratio inspirals}},
{\em Phys. Rev. D} {\bf 95}, 103012 (2017),
[{{\tt arXiv:1703.09722}}].






\bibitem{Huerta:2011zi}
E.~A.~Huerta, J.~R.~Gair, and D.~A.~Brown,
{\it {Importance of including small body spin effects in the modelling of intermediate mass-ratio inspirals. II Accurate parameter extraction of strong sources using higher-order spin effects}},
{\em Phys. Rev. D} {\bf 85}, 064023 (2012),
[{{\tt arXiv:1111.3243}}].

\bibitem{Piovano:2021iwv}
G.~A.~Piovano, R.~Brito, A.~Maselli, and P.~Pani,
{\it {Assessing the detectability of the secondary spin in extreme mass-ratio inspirals with fully relativistic numerical waveforms}},
{\em Phys. Rev. D} {\bf 104}, 124019 (2021),
[{{\tt arXiv:2105.07083}}].


\bibitem{Warburton:2017sxk}
N.~Warburton, T.~Osburn, and C.~R.~Evans,
{\it {Evolution of small-mass-ratio binaries with a spinning secondary}},
{\em Phys. Rev. D} {\bf 96}, 084057 (2017),
[{{\tt arXiv:1708.03720}}].




\bibitem{Skoupy:2021asz}
V.~Skoup\'y and G.~Lukes-Gerakopoulos,
{\it {Spinning test body orbiting around a Kerr black hole: Eccentric equatorial orbits and their asymptotic gravitational-wave fluxes}},
{\em Phys. Rev. D} {\bf 103}, 104045 (2021),
[{{\tt arXiv:2102.04819}}].

\bibitem{Mathews:2021rod}
J.~Mathews, A.~Pound, and B.~Wardell,
{\it {Self-force calculations with a spinning secondary}},
{\em Phys. Rev. D} {\bf 105}, 084031 (2022),
[{{\tt arXiv:2112.13069}}].

\bibitem{Piovano:2020zin}
G.~A.~Piovano, A.~Maselli, and P.~Pani,
{\it {Extreme mass ratio inspirals with spinning secondary: a detailed study of equatorial circular motion}},
{\em Phys. Rev. D} {\bf 102}, 024041 (2020),
[{{\tt arXiv:2004.02654}}].

\bibitem{Drummond:2023wqc}
L.~V.~Drummond, P.~Lynch, A.~G.~Hanselman, D.~R.~Becker, and S.~A.~Hughes,
{\it {Extreme mass-ratio inspiral and waveforms for a spinning body into a Kerr black hole via osculating geodesics and near-identity transformations}},
{\em Phys. Rev. D} {\bf 109}, 064030 (2024),
[{{\tt arXiv:2310.08438}}].


\bibitem{Wald:1972sz}
R.~M.~Wald,
{\it {Gravitational spin interaction}},
{\em Phys. Rev. D} {\bf 6}, 406 (1972).

\bibitem{Hanson:1974qy}
A.~J.~Hanson and T.~Regge,
{\it {The Relativistic Spherical Top}},
{\em Annals Phys.} {\bf 87}, 498 (1974).


\bibitem{Mathisson:1937ms}
M.~Mathisson,
{\it {Neue mechanik materieller systemes}},
{ Acta Phys. Polon.} {\bf 6} (1937) 163.



\bibitem{Papapetrou:1951gr}
A.~Papapetrou,
{\it {Spinning test particles in general relativity. 1.}},
{ Proc. Roy. Soc. Lond.} {\bf A 209} (1951) 248.

\bibitem{Corinaldesi:1951gr}
E.~Corinaldesi and A.~Papapetrou,
{\it {Spinning test particles in general relativity. 2.}},
{ Proc. Roy. Soc. Lond.} {\bf A 209} (1951) 259.


\bibitem{Tulczyjew:1959tr}
W.~Tulczyjew,
{\it Motion of multipole particles in general relativity theory},
{ Acta Phys. Pol} {\bf 18} (1959) 94.

\bibitem{Dixon:1964gr}
W.~G. Dixon,
{\it A covariant multipole formalism for extended test bodies in general relativity},
{ Il Nuovo Cimento (1955-1965)} {\bf 34} (1964) 317.


\bibitem{Suzuki:1998vy}
S.~Suzuki and K.-i. Maeda,
{\it {Innermost stable circular orbit of a spinning particle in Kerr space-time}},
{\em Phys. Rev. D} {\bf 58}, 023005 (1998),
[{{\tt arXiv:gr-qc/9712095}}].




\bibitem{Han:2008zzf}
W.-B.~Han,
{\it {Chaos and dynamics of spinning particles in Kerr spacetime}},
{\em Gen. Rel. Grav.} {\bf 40}, 1831 (2008),
[{{\tt arXiv:1006.2229}}].

\bibitem{Jefremov:2015gza}
P.~I. Jefremov, O.~Y. Tsupko, and G.~S. Bisnovatyi-Kogan,
{\it {Innermost stable circular orbits of spinning test particles in Schwarzschild and Kerr space-times}},
{\em Phys. Rev. D} {\bf 91}, 124030 (2015),
[{{\tt arXiv:1503.07060}}].




\bibitem{Harms:2016ctx}
E.~Harms, G.~Lukes-Gerakopoulos, S.~Bernuzzi, and A.~Nagar,
{\it {Spinning test body orbiting around a Schwarzschild black hole: Circular dynamics and gravitational-wave fluxes}},
{\em Phys. Rev. D} {\bf 94}, 104010 (2016),
[{{\tt arXiv:1609.00356}}].

\bibitem{LukesGerakopoulos:2017cru}
G.~Lukes-Gerakopoulos, E.~Harms, S.~Bernuzzi, and A.~Nagar,
{\it {Spinning test-body orbiting around a Kerr black hole: circular dynamics and gravitational-wave fluxes}},
{\em Phys. Rev. D} {\bf 96}, 064051 (2017),
[{{\tt arXiv:1707.07537}}].

\bibitem{Zhang:2017nhl}
Y.-P.~Zhang, S.-W.~Wei, W.-D.~Guo, T.-T.~Sui, and Y.-X.~Liu,
{\it {Innermost stable circular orbit of spinning particle in charged spinning black hole background}},
[{{\tt arXiv:1711.09361}}].

\bibitem{Mukherjee:2018zug}
S.~Mukherjee and K.RajeshNayak,
{\it {Off-equatorial stable circular orbits for spinning particles}},
{\em Phys. Rev. D} {\bf 98}, 084023 (2018),
[{{\tt arXiv:1804.06070}}].





\bibitem{Zhang:2018omr}
Y.-P.~Zhang, S.-W.~Wei, P.~Amaro-Seoane, J.~Yang, and Y.-X.~Liu,
{\it {Motion deviation of test body induced by spin and cosmological constant in extreme mass ratio inspiral binary system}},
{\em Eur.Phys.J. C} {\bf  79}, 856 (2019),
[{{\tt arXiv:1812.06345}}].

\bibitem{Zhang:2019oet}
M.~Zhang and W.-B.~Liu,
{\it {Innermost stable circular orbits of charged spinning test particles}},
{\em Phys. Lett. B} {\bf 789}, 393 (2019),
[{{\tt arXiv:1812.10115}}].

\bibitem{Antoniou:2019awm}
I.~Antoniou, D.~Papadopoulos, and L.~Perivolaropoulos,
{\it {Spinning particle orbits around a black hole in an expanding background}},
{\em Class. Quant. Grav.} {\bf 36}, 085002 (2019),
[{{\tt arXiv:1903.03835}}].

\bibitem{Nucamendi:2019qsn}
U.~Nucamendi, R.~Becerril, and P.~Sheoran,
{\it {Bounds on spinning particles in their innermost stable circular orbits around rotating braneworld black hole}},
{\em Eur. Phys. J. C} {\bf 80}, 35 (2020),
[{{\tt arXiv:1910.00156}}].

\bibitem{Zhang:2020qam}
Y.-P.~Zhang, S.-W.~Wei, and Y.-X.~Liu,
{\it {Spinning Test Particle in Four-Dimensional Einstein--Gauss--Bonnet Black Holes}},
{\em Universe} {\bf 6}, 103 (2020),
[{{\tt arXiv:2003.10960}}].






\bibitem{Semerak:2015dza}
O.Semer\'ak and M.\v{S}r\'amek,
{\it {Spinning particles in vacuum spacetimes of different curvature types}},
{\em Phys. Rev. D} {\bf 92}, 064032 (2015),
[{{\tt arXiv:1505.01069}}].

\bibitem{Mukherjee:2018dmm}
S.~Mukherjee,
{\it {Periastron shift for a spinning test particle around naked singularities}},
{\em Phys. Rev. D} {\bf 97}, 124006 (2018).

\bibitem{Toshmatov:2019bda}
B.~Toshmatov and D.~Malafarina,
{\it {Spinning test particles in the $\gamma$ spacetime}},
{\em Phys. Rev. D} {\bf 100}, 104052 (2019),
[{{\tt arXiv:1910.11565}}].

\bibitem{Benavides-Gallego:2021lqn}
C.~A.~Benavides-Gallego, W.-B.~Han, D.~Malafarina, B.~Ahmedov, and A.~Abdujabbarov,
{\it {Spinning test particle motion around a traversable wormhole}},
{\em Phys. Rev. D} {\bf 104}, 084024 (2021),
[{{\tt arXiv:2107.07998}}].


\bibitem{Yang:2022jno}
K.~Yang, B.-M.~Gu, and Y.-P.~Zhang,
{\it {Motion of spinning particles around electrically charged black hole in Eddington-inspired Born--Infeld gravity}},
{\em Eur. Phys. J. C} {\bf 82}, 293 (2022),
[{{\tt arXiv:2111.00864}}].


\bibitem{Zhang:2022qzw}
Y.-P.~Zhang, Y.-B.~Zeng, Y.-Q.~Wang, S.-W.~Wei, and Y.-X.~Liu,
{\it {Equatorial orbits of spinning test particles in rotating boson stars}},
{\em Eur. Phys. J. C} {\bf 82}, 809 (2022),
[{{\tt arXiv:2201.01498}}].

\bibitem{Chen:2024sbc}
K.~Chen and S.-W.~Wei,
{\it {Motion of spinning particles around a polymer black hole in loop quantum gravity}},
{\em Phys. Rev. D} {\bf 110}, 024041 (2024),
[{{\tt arXiv:2403.14164}}].

\bibitem{Witzany:2023bmq}
V.~Witzany and G.~A.~Piovano,
{\it {Analytic Solutions for the Motion of Spinning Particles near Spherically Symmetric Black Holes and Exotic Compact Objects}},
{\em Phys. Rev. Lett.} {\bf 132}, 171401 (2024),
[{{\tt arXiv:2308.00021}}].

\bibitem{Liu:2024lda}
Y.-L.~Liu and X.-D.~Zhang,
{\it {Analytic solutions for the motion of spinning particles near brane-world black hole}},
[{{\tt arXiv:2408.06852}}].



\bibitem{Costa:2014nta}
L.~F.~O.~Costa and J.~Nat\'ario,
{\it {Center of mass, spin supplementary conditions, and the momentum of spinning particles}},
{\em Fund. Theor. Phys.} {\bf 179}, 215 (2015),
[{{\tt arXiv:1410.6443}}].


\bibitem{Wei:2019zdf}
S.-W.~Wei, J.~Yang, and Y.-X.~Liu,
{\it {Geodesics and periodic orbits in Kehagias-Sfetsos black holes in deformed Ho\v{r}ava-Lifshitz gravity}},
{\em Phys. Rev. D}  {\bf 99}, 104016  (2019),
[{{\tt  arXiv:1904.03129}}].




\bibitem{Deriglazov:2015wde}
A.~A.~Deriglazov and W.~G.~Ram\'\i{}rez,
{\it {Ultrarelativistic Spinning Particle and a Rotating Body in External Fields}},
{\em Adv. High Energy Phys.} {\bf 2016}, 1376016 (2016),
[{{\tt arXiv:1511.00645}}].

\bibitem{Deriglazov:2017jub}
A.~A.~Deriglazov and W.~G.~Ram\'\i{}rez,
{\it {Recent progress on the description of relativistic spin: vector model of spinning particle and rotating body with gravimagnetic moment in General Relativity}},
{\em Adv. Math. Phys.} {\bf 2017}, 7397159 (2017),
[{{\tt arXiv:1710.07135}}].




\bibitem{Levin:2008mq}
J.~Levin and G.~Perez-Giz,
{\it {A Periodic Table for Black Hole Orbits}},
{\em Phys. Rev. D}  {\bf 77}, 103005  (2008)
[{{\tt arXiv:0802.0459}}].

























\end{thebibliography}
\end{document}